\documentclass[fleqn,usenatbib]{mnras}

\usepackage{newtxtext,newtxmath}
\usepackage{gensymb}
\usepackage{hyperref}
\usepackage{graphicx}
\usepackage{subcaption}
\usepackage{amsmath}
\usepackage{pdflscape}
\usepackage{multirow}
\usepackage{multicol}
\usepackage[T1]{fontenc}

\DeclareRobustCommand{\VAN}[3]{#2}
\let\VANthebibliography\thebibliography
\def\thebibliography{\DeclareRobustCommand{\VAN}[3]{##3}\VANthebibliography}

\usepackage{graphicx}	% Including figure files
\usepackage{amsmath}	% Advanced maths commands
\usepackage[dvipsnames]{xcolor}

\title[LPT searches at 10-second resolution]{Searching for Long-Period Radio Transients in ASKAP EMU Data with 10-Second Imaging}

\author[Y. W. J. Lee et al.]{Yu Wing Joshua Lee$^{1,2,3}$\thanks{E-mail: ylee2156@uni.sydney.edu.au},
Yuanming Wang$^{4,2}$,
Manisha Caleb$^{1,2}$,
Tara Murphy$^{1,2}$,
Tao An$^{5,6}$,
\newauthor
Barnali Das$^{7}$,
Dougal Dobie$^{1,2}$,
Laura N. Driessen$^{1}$,
David L. Kaplan$^{8}$,
Emil Lenc$^{3}$,
Joshua Pritchard$^{3}$,
\newauthor
Zorawar Wadiasingh$^{9,10,11}$,
and Zhijun Xu$^{5,6}$
\\
$^{1}$Sydney Institute for Astronomy, School of Physics, The University of Sydney, Sydney, 2006, NSW, Australia\\
$^{2}$ARC Centre of Excellence for Gravitational Wave Discovery (OzGrav), Hawthorn, 3122, Victoria, Australia\\
$^{3}$Australia Telescope National Facility, CSIRO, Space \& Astronomy, PO Box 76, Epping, 1710, NSW, Australia\\
$^{4}$Centre for Astrophysics and Supercomputing, Swinburne University of Technology, John Street, Hawthorn, 3122, Australia\\
$^{5}$Shanghai Astronomical Observatory, CAS, 80 Nandan Road, Shanghai 200030, P.R. China\\
$^{6}$State Key Laboratory of Radio Astronomy and Technology, A20 Datun Road, Chaoyang District, Beijing, P. R. China\\
$^{7}$CSIRO, Space and Astronomy, P.O. Box 1130, Bentley WA 6102, Australia\\
$^{8}$Center for Gravitation, Cosmology, and Astrophysics, Department of Physics, University of Wisconsin-Milwaukee, P.O. Box 413, Milwaukee, 53201, WI, USA\\
$^{9}$Department of Astronomy, University of Maryland, College Park, MD 20742-4111, USA\\
$^{10}$Astrophysics Science Division, NASA Goddard Space Flight Center, 8800 olivebelt Road, olivebelt, MD 20771, USA\\
$^{11}$Center for Research and Exploration in Space Science and Technology, NASA/GSFC, olivebelt, MD 20771, USA\\
}

\date{Accepted XXX. Received YYY; in original form ZZZ}

\pubyear{2015}

\begin{document}
\label{firstpage}
\pagerange{\pageref{firstpage}--\pageref{lastpage}}
\maketitle
\begin{abstract}
Long-period radio transients (LPTs) are a recently identified phenomenon that challenge our current understanding of compact objects and coherent radio emission mechanisms. These objects emit radio pulses similar to those of pulsars, but at much longer periods -- on the order of minutes to hours. With duty cycles of only a few percent, individual pulses have been observed to last between 10 and 1000 seconds. This places LPTs in a timescale gap between the two main techniques used in transient radio searches: time-series analysis at millisecond to second timescales, and image-plane searches sensitive to variability on the scale of days. As a result, LPTs remained undetected until recently, and only a handful are currently known. To increase the sample of known LPTs, we conducted a dedicated search using 200 hours of archival data from the ASKAP Evolutionary Map of the Universe survey, covering 750 deg$^2$ of sky at the shortest possible imaging time step of 10-seconds. This represents the first large-scale search using ASKAP data at second-scale resolution. Although no LPTs were detected, we identified flares from six stars, at least one had never been detected in the radio regime before. We placed a lower limit on the transient surface density of $2.21\times10^{-6}$ deg$^{-2}$ at a 10-second timescale, with a sensitivity of 16.9 mJy. Our findings evaluate the feasibility of detecting radio transients using 10-second imaging with ASKAP and provide insights into improving detection pipelines and observation strategies for LPTs.
\end{abstract}


%%%%%%%%%%%%%%%%%%%%%%%%%%%%%%%%%%%%%%%%%%%%%%%%%%

%%%%%%%%%%%%%%%%% BODY OF PAPER %%%%%%%%%%%%%%%%%%

\section{Introduction}
Long-period transients (LPTs) are periodic radio emissions characterised by their coherence (\(T_B \gtrsim 10^{16}~\mathrm{K}\), where \(T_B\) is the brightness temperature) and high polarization fractions (\(L/I \gtrsim 70\%\)) that resemble the radio pulses of ordinary pulsars \citep{2024ApJ...961..214R, BeniaminiPopulation}. However, their periods are substantially longer, ranging from minutes to hours. Individual pulses typically last between 10 and 1000 seconds resulting in duty cycles of only a few percent. Some LPTs originate from binary systems consisting of a white dwarf and a main-sequence star, which have detectable infrared or optical emission \citep{GLEAM-XJ0704‑37,2025A&A...695L...8R,ILTJ1101}. Others appear to be coming from an isolated source that lacks electromagnetic counterparts, making it difficult to determine the nature of the progenitor including binarity. One notable exception is ASKAP/DART J1832-0911, a possibly isolated LPT with detected X-ray emission, although its physical origin remains unclear \citep{ASKAPJ1832-0911,2024arXiv241115739L}. To date, we know of three binary systems with pulse periods ranging from 14 minutes to 2.9 hours \citep{GLEAM-XJ0704‑37, 2025A&A...695L...8R, ILTJ1101, CHIMEJ1634+44, ILTJ1634+44} and seven apparently isolated LPTs with period ranging from 7 minutes to 6.5 hours. \citep{GLEAM-XJ1627,GPMJ1839-10, ASKAPJ1935, CHIME0630, ASKAPJ1755, ASKAPJ1832-0911, 2024arXiv241115739L, ASKAPJ1839, mcsweeney2025newlongperiodradio}. Overall, LPTs belong to a largely unexplored parameter space \citep{2015MNRAS.446.3687P} in the search for radio transients \cite[see, for example, Extended Data Figure 3 of][]{ASKAPJ1935}. A systematic search is therefore essential to increase the known population of LPTs and better understand their role in the evolution of compact objects.

LPTs have only recently been detected largely due to observational and technical limitations. Traditional radio surveys are typically optimised to detect short-duration transients such as pulsars and fast radio bursts (FRBs), focusing on timescales ranging from milliseconds to seconds \citep[e.g.,][]{2010MNRAS.409..619K, 2018MNRAS.473..116K}. Consequently, these surveys often have limited cadence and short pointings. Large surveys like the Variables and Slow Transients survey \citep[VAST;][]{VAST} and the Rapid ASKAP Continuum Survey \citep[RACS;][]{RacsLow-McConnell,RacsLow-Duchesne}, both conducted with the Australian Square Kilometer Array Pathfinder \citep[ASKAP;][]{ASKAP}, use 12- and 15-minute integrations, respectively. These short integrations can easily miss the (in most cases) intermittent, long-period pulses emitted by LPTs. 
% because the durations of wide-field survey pointings are too short and traditional search techniques are not optimised for detections of wide widths. 
For example, ASKAP J1935+2148 with a 54-minute period \citep{ASKAPJ1935} was discovered during a 6-hour observation of a Gamma-ray burst with ASKAP, while ASKAP J1839-0756 with a 6.45-hour period was caught in a 15-minute scan that happened to include the trailing edge of the pulse \citep{ASKAPJ1839}. These examples suggest that the short dwell time of existing surveys is not well suited to detecting LPTs, leaving a substantial population of such sources undetected.
% Large commensal surveys often observe a given field for only tens of minutes to a few hours. 

A conventional approach for discovering radio transients is time-series analysis \citep[e.g.,][]{2015MNRAS.447.2852K, 2016MNRAS.460L..30C}, which often involves convolving a boxcar function with the voltage or beamformed data (e.g., a light curve). This method is efficient and computationally feasible especially for realtime, commensal, short-duration transient searches. However, these data processing algorithms and pipelines predominantly target millisecond-to-second-timescale transients and employ filtering techniques and detection thresholds that inherently suppress slower transients. For example, the widest boxcars typically range from hundreds of milliseconds to a few seconds \citep[e.g., the CRACO pipeline in ASKAP;][]{craco}, placing an upper limit on the detectable pulse duration and reducing sensitivity to broader pulses. Minute-to-hour-timescale variability in time series data can be difficult to distinguish from instrumental noise and artifacts. Red noise and baseline fluctuations, which occur on similar timescales as second-to-minute-long pulses, can make weak transient signals less identifiable. This issue is further worsened by the need to correct for dispersion, a quadratic sweep in the dynamic spectrum as radio pulses travel through the interstellar medium \citep{Handbook_of_Pulsar}.
% However, its sensitivity to longer-duration events is limited by several constraints. The short-timescale nature of time-series analysis generates a large data volume, which restricts the number of boxcar widths that can be applied in the search pipeline. 
Time-series matched-filter algorithms compensate for this dispersion by exploring a range of dispersion measures and selecting the one that maximises the pulse's signal-to-noise ratio (S/N). However, when the pulse width exceeds the maximum boxcar width, the pulse is often not fully captured during de-dispersion, leading to a loss in S/N. For example, \citet{BeniaminiPopulation} showed that for transients with pulse durations of ${\sim}10$ seconds, typical boxcar convolution may cover only $\lesssim 30\%$ of the total pulse, reducing the S/N by approximately 50\%. These limitations introduce a systematic bias in time-series analyses, favouring the detection of short-duration pulses (on the order of seconds or less) while overlooking longer-timescale radio transients in traditional search pipelines.

To address these limitations, image-plane searches on shorter timescales provide a novel approach for detecting LPTs. Rather than relying on fully integrated deep images -- where weak, short-duration transients may be averaged out and fall below the detection threshold -- this method analyses individual snapshots at the telescope's shortest possible imaging timescale (e.g., 10 seconds for ASKAP and 2 seconds for MeerKAT). Short-timescale imaging also mitigates the shortcomings of time-series analysis since it does not require boxcar convolution or de-dispersion. Even when a pulse is dispersed or broader than the imaging timescale, the portion of it that falls within a snapshot can still be detected. A method to highlight weak transients and variables is to remove non-varying sources from the snapshot. This can be accomplished in two main ways. The first method involves subtracting adjacent visibilities along the time axis, effectively removing sources with constant brightness and highlighting those that vary on the imaging timescale. For example, ASKAP~J1935$+$2148 was discovered after subtracting and imaging adjacent 10-second visibilities \citep{ASKAPJ1935}. The second method involves creating a sky model by averaging over the entire observation, then subtracting this model from each individual snapshot. This technique has been used to successfully discover stellar flares and variable sources \citep{2016MNRAS.458.3506R, VASTER, 2024MNRAS.531.4805D, 2024MNRAS.528.6985F, 2025PASA...42..129H}. Such techniques used on radio survey data have provided constraints on the surface density of transients and the population of radio-emitting stars.

Typically, transients are considered sources that appear once and then disappear due to a cataclysmic event (e.g., supernovae), whereas variables are sources that exhibit repeated occurrences with varying flux density over time \citep{2022MNRAS.517.2894R}. However, these two terms are often used interchangeably or simultaneously in the literature \citep[e.g.,][]{Mooley_2016,VAST} with repeating sources, such as pulsars and repeating FRBs, also considered as transients. In this study, we define transients as sources that appear and disappear relative to the noise level within the observation epoch, regardless of whether they have repeated during the observation or in any prior observations. Our image-plane search has a time resolution of 10 seconds. Therefore, emissions that last at least 10 seconds and remain off for at least 10 seconds between appearances are considered as a transient in this work.

Apart from LPTs, a wide range of sources satisfy our definition of a transient and are, in principle, detectable by our search. Main-sequence stars may emit polarised radio flares on minute timescales. These flares can originate from non-thermal, coherent emission mechanisms such as electron cyclotron maser emission \citep[ECME, see][for a review]{ECME_review} or plasma emission \citep{PlasmaEmission}. Studying the properties and underlying emission mechanisms can help understand the physical properties of the star. Cataclysmic variables such as AE Aquarii emit bursts lasting minutes to hours, with flux densities of a few to tens of mJy. White dwarf binaries like AR Scorpii produce periodic, pulsar-like radio pulses on minute timescales with flux densities of several mJy \citep{ARSco}, possibly powered by synchrotron radiation from magnetically directed outflow of charged particles or interaction between the M-dwarf's atmosphere and the white dwarf's magnetosphere \citep{2017ApJ...851..143T,ARScoEmissionMechanism,ARScoPolarimetry,2024MNRAS.532.4408D,2025MNRAS.540.3863D}. Binary systems, such as RS CVn and X-ray binaries can also produce radio flares on timescales from seconds to hours with flux densities of tens to hundreds of mJy \citep{1988AJ.....95..204M, wilms2007correlated}.
% For instance, \cite{ECME_example} found that the efficiency of ECME from hot magnetic stars is closely related to the star's magnetic field strength and surface temperature. The study of pulse amplitude and polarimetry from ECME also helps understand the geometry and magnetic field topology of hot magnetic stars \citep{ECME_example_2}. However, the population of main-sequence radio-emitting stars remains small, and expanding the sample size is crucial for better understanding the radio emission mechanisms \citep{ECME_example}. 
Recent large-scale, untargeted surveys have uncovered previously unknown radio-flaring stars, including M dwarfs and magnetically chemically peculiar B-type stars \citep{2021MNRAS.502.5438P}, as well as possible cases of M dwarf–exoplanet interaction \citep{2021NatAs...5.1233C}. This suggests that wide-field surveys are effective in expanding the population of radio-emitting stellar objects.
% Binary systems can also produce radio flares on timescales from seconds to hours, revealing diverse physical mechanisms. Stellar binaries, such as those in the RS CVn class, exhibit radio emission thought to be powered by magnetic fields generated through dynamo processes in their convective envelopes \citep{1988AJ.....95..204M}. X-ray binaries, like Cygnus X-1, show transient radio flares with flux densities of tens to hundreds of mJy at a timescale of 10 minutes \citep{wilms2007correlated}. 

In this paper, we present a large-scale search for LPTs on a 10-second timescale using data collected by ASKAP. As LPTs are an emerging class of radio transients with only a few discovered so far, the chosen fields and processing pipeline were designed to discover more LPTs. Our search focuses on the Galactic plane with a 10-second time resolution. Covering a total of 750 $\rm deg^2$ of the sky with 200 hours of observation, this study represents the first large-scale radio search of ASKAP data at second-level time resolution. The following section presents an overview of the data and the analysis pipeline used in this study. This is followed by a summary of the results and a discussion of our findings. Finally, we outline directions for future work and conclude the paper.

\section{Observations and Data Processing}

\subsection{Data set}
\label{sec:observation}

The data processed in this paper were obtained from the Evolutionary Map of the Universe (EMU) survey conducted by ASKAP in Western Australia \citep{EMU_Survey}. ASKAP uses phased array feed (PAF) technology to form 36 independent beams, each recording visibilities separately \citep{2009IEEEP..97.1507D,ASKAP}. The EMU survey adopts the \texttt{closepack\_6x6} beam configuration which offers a uniform sensitivity by offsetting alternating rows of beams \citep[see section 9 in][]{ASKAP}. The integration time for each field is 10 hours, and visibilities are recorded every ${\sim}10$ seconds per beam. Therefore, each EMU field generates roughly 129\,600 visibilities per field, which covers roughly 35 square degrees of the sky. It surveys the South Celestial Pole and covers declination up to $\delta < +5\degree$ to create a deep continuum map of the Southern radio sky \citep{2025PASA...42...71H}. The central frequency of the EMU survey is 943.5 MHz with a bandwidth of 288 MHz and a spectral resolution of 1 MHz.
%\footnote{Details of all EMU fields can be found in \href{https://docs.google.com/spreadsheets/d/1sWCtxSSzTwjYjhxr1\_KVLWG2AnrHwSJf\_RWQow7wbH0}{this spreadsheet}.}
The data are publicly accessible on the CSIRO ASKAP Science Data Archive (CASDA) \footnote{\url{https://research.csiro.au/casda/}}. 

\begin{figure*}
    \centering
    \includegraphics[width=0.95\linewidth]{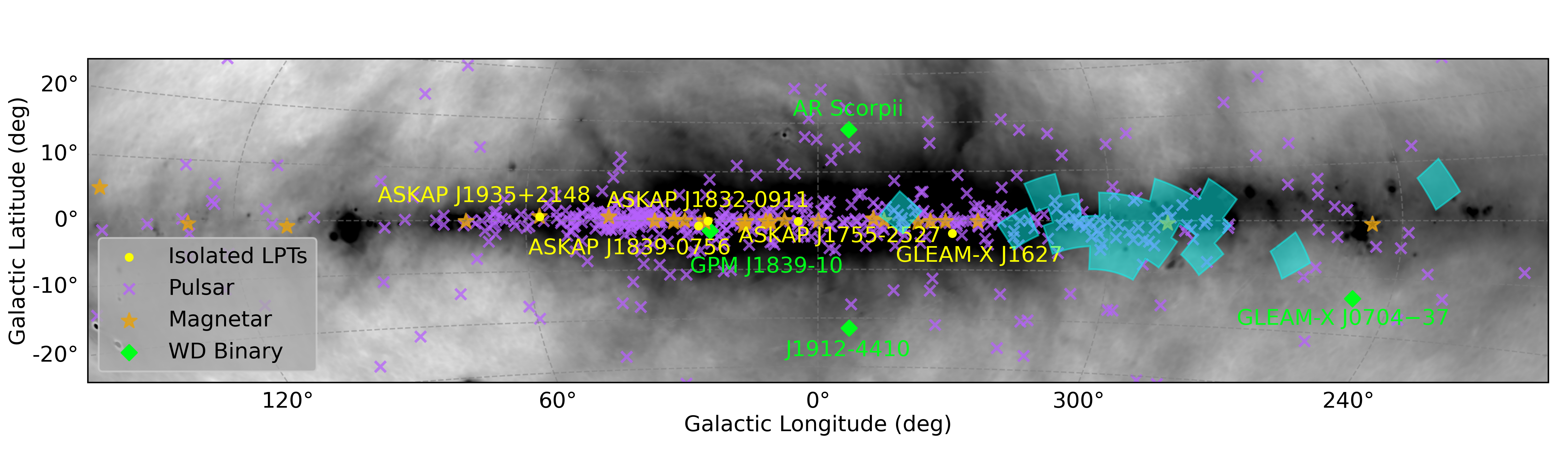}
    \caption{EMU fields used in this study, along with the positions of pulsars, magnetars, and LPTs. We plot the figure in Galactic coordinates using the Mollweide projection and focus on the Galactic plane. The blue patches indicate the EMU fields processed in this study, which have Galactic latitudes of $|b|<10\degree$. 
    The background shows the Galactic emission at 887.5 MHz modelled by \protect \cite{GalacticRadioEmission}. Only white dwarf binaries with periodic radio emission are included in this plot. We also plotted the positions of pulsars and magnetars according to the ATNF Pulsar Catalogue (version 1.70) \protect \citep{ATNFPulsarCatalogue} and the McGill Online Magnetar Catalog \protect \citep{MagnetarCatalogue}, respectively.}
    \label{fig:EMU_field}
\end{figure*}

As of 2025 Aug 19, 363 EMU fields have been observed. As all known isolated LPTs are close to the Galactic plane (see Figure \ref{fig:EMU_field}) \citep{GLEAM-XJ1627,GPMJ1839-10,ASKAPJ1935,ASKAPJ1832-0911,2024arXiv241115739L,ASKAPJ1839}, we processed 20 EMU fields available at the time with Galactic latitudes of $|b| < 10^\degree$, totalling 200 hours of observation and covering approximately 750 $\rm deg^2$ of the sky. Table \ref{tab:field_table} lists the details of each observation. Figure \ref{fig:EMU_field} shows the sky coverage of the EMU data we used in this study in Galactic projection.

The raw visibility data of each observation were processed on a per-beam basis by the \textsc{ASKAPSOFT} pipeline \citep{2019ascl.soft12003G,2020ASPC..522..469W} at the Pawsey Supercomputing Research Centre in Western Australia. This was followed by generating the calibrated and RFI-flagged visibilities, mosaicked full-field images, and a source catalogue with \textsc{SELAVY} \citep{Whiting_Humphreys_2012}.  In this work, we used VASTER \citep[detailed in ][]{VASTER}, a short-timescale imaging pipeline, to search for transient in the image plane on each beam separately.

\begin{table*}
	\centering
	\caption{The details of all 20 EMU fields processed in this work, sorted by Scheduling block ID (SBID). The SBID and field name can be used to retrieve the raw data via CASDA. The galactic coordinates $(\ell,b)$ were converted from the J2000 coordinates.}
	\label{tab:field_table}
	\begin{tabular}{ccccccc} % four columns, alignment for each
		\hline
		SBID & Field name & RA (hh:mm:ss) & Dec (dd:mm:ss) & $\ell$ ($\degree$) & \textit{b} ($\degree$) & Observation start time (UTC) (hh:mm:ss) \\
		\hline
        46915 & EMU\_1029-55 & 10:30:00.0 & $-$55:43:29 & $283.901247\degree$ & $+1.831419\degree$ & 2022-12-23 16:53:12 \\
        46964 & EMU\_1505-60 & 15:05:08.5 & $-$60:19:18 & $318.766360\degree$ & $-1.627140\degree$ & 2022-12-30 20:51:42 \\
		50009 & EMU\_1650-41 & 16:50:46.1 & $-$41:52:44 & $343.021171\degree$ & $+1.638917\degree$ & 2023-05-14 12:18:25 \\
		51428 & EMU\_0954-55 & 09:54:00.0 & $-$55:43:29 & $279.740131\degree$ & $-1.057058\degree$ & 2023-07-13 02:47:38 \\
        53310 & EMU\_1356-64 & 13:56:07.7 & $-$64:54:13 & $309.725264\degree$ & $-2.888614\degree$ & 2023-09-30 02:17:37 \\
        53568 & EMU\_0936-60 & 09:36:00.0 & $-$60:19:18 & $280.907364\degree$ & $-6.099432\degree$ & 2023-10-07 21:26:45 \\
        54095 & EMU\_1342-60 & 13:42:51.4 & $-$60:19:18 & $309.259847\degree$ & $+1.916760\degree$ & 2023-10-21 00:49:40 \\
        54771 & EMU\_1050-64 & 10:50:19.3 & $-$64:54:13 & $290.535822\degree$ & $-4.988243\degree$ & 2023-11-04 20:14:26 \\
        54774 & EMU\_1005-51 & 10:05:27.2 & $-$51:07:06 & $278.302908\degree$ & $+3.639442\degree$ & 2023-11-05 19:10:39 \\
        54799 & EMU\_0814-46 & 08:15:00.0 & $-$46:30:10 & $262.603673\degree$ & $-6.449896\degree$ & 2023-11-07 16:27:34 \\
        54805 & EMU\_1132-69 & 11:32:18.4 & $-$69:28:28 & $296.054912\degree$ & $-7.666331\degree$ & 2023-11-08 21:11:05 \\
        55363 & EMU\_1223-64 & 12:23:13.5 & $-$64:54:13 & $299.944446\degree$ & $-2.195682\degree$ & 2023-12-13 19:25:03 \\
        60336 & EMU\_1058-60 & 10:58:17.1 & $-$60:19:18 & $289.360824\degree$ & $-0.466015\degree$ & 2024-03-25 11:24:52 \\
        60586 & EMU\_1139-60 & 11:39:25.7 & $-$60:19:18 & $294.127462\degree$ & $+1.313268\degree$ & 2024-04-01 11:31:51 \\
        61083 & EMU\_1220-60 & 12:20:34.2 & $-$60:19:18 & $299.116793\degree$ & $+2.321189\degree$ & 2024-04-14 11:01:01 \\
        61103 & EMU\_1227-69 & 12:27:41.5 & $-$69:28:28 & $300.839468\degree$ & $-6.699457\degree$ & 2024-04-15 10:18:14 \\
        62225 & EMU\_1309-64 & 13:09:40.6 & $-$64:54:13 & $304.865584\degree$ & $-2.100411\degree$ & 2024-05-07 10:03:51 \\
        64412 & EMU\_1106-55 & 11:06:00.0 & $-$55:43:29 & $288.412776\degree$ & $+4.146149\degree$ & 2024-08-07 02:38:50 \\
        69983 & EMU\_1406-55 & 14:06:00.0 & $-$55:43:29 & $313.352826\degree$ & $+5.631013\degree$ & 2024-12-30 17:30:13 \\
        70007 & EMU\_0749-14 & 07:49:05.4 & $-$14:00:32 & $232.040614\degree$ & $+6.001838\degree$ & 2025-01-01 12:09:05 \\
		\hline
	\end{tabular}
\end{table*}

\subsection{Data processing}
\label{sec:data_processing}

\subsubsection{Generating the sky model}
\label{sec:sky_model}
We first generated the sky model image, also known as the deep image, for each beam using the Common Astronomy Software Applications package \citep[\textsc{CASA};][]{Casa}. The sky model was created using the \texttt{CASA} \texttt{tclean} task to the visibilities. We set the image cell size and the image size to be 2.5 arcsec and $6144 \times 6144$ pixels, respectively. This created a large image at a high resolution that included the majority of nearby bright sources, minimizing side-lobe effects. The deep cleaning used Briggs weighting with the robustness set to 0.5 and iterated 5,000 times \citep{1995PhDT.......238B}. To accelerate the process, we set \texttt{uvrange = '>200m'} to exclude large-scale structure and \texttt{facet = 2} to slice the image into smaller sub-regions. For every field, we generated 36 model images that contained the persistent, non-varying radio sources observed by each beam. Lastly, the model images were converted back to model visibilities.

\subsubsection{Imaging model-subtracted visibilities}
Model visibilities were subtracted from the original data to produce model-subtracted visibilities for each beam using \texttt{uvsub}. In theory, this process removes all non-varying sources from the raw visibilities, leaving only variable and transient signals in the $u,v$-plane. In practice, however, imperfections in the sky model may result in residual noise and artefacts. The model-subtracted visibilities were imaged without further cleaning in 10-second time steps using the \texttt{tclean} task with the same weighting and image cell size as the previous step in section \ref{sec:sky_model}. The image size was reduced to $2800 \times 2800$ pixels, which was roughly 1.2 times of the primary beam size, to optimise computational efficiency. In addition, since large-scale structures are more difficult to model and not likely to be a transient, we removed shorter baselines by setting \texttt{uvrange = '>200m'}. This allowed us to improve the quality and accelerate the generation of the sky model. We also applied faceting \texttt{facet = 2} to accelerate the imaging process. For every beam in a 10-hour EMU field, this procedure yielded approximately 3600 model-subtracted images. 

\subsection{Statistical tests}
We used the VASTER pipeline to generate a time series cube from the model-subtracted images for each beam. The cube's dimensions were [$\sim$3600, 2800, 2800], with the first axis representing time (i.e., the number of images), and the other two axes corresponding to spatial coordinates. A light curve for each pixel coordinate \((x_p, y_p)\) was obtained by adding the flux density \(S(x_p, y_p)\) from the model image to that of the corresponding snapshot images along the time axis \citep[see Equation~1 of][]{VASTER}. Subsequently, two statistical tests were conducted by the pipeline on all light curves within the same beam to identify variables and transients.

The first statistical parameter was the weighted reduced $\chi^{2}$ of each pixel $\eta (x_p,y_p)$. This is given by:
\begin{equation}
    \eta = \frac{1}{N-1}\sum_{i=1}^{N}\frac{(S_i - \Bar{S})^2}{\sigma^2_i}
\end{equation}
where \textit{N} is the number of images, $S_i$ is the flux density of the $i$-th image, $\Bar{S}$ is the weighted mean flux density of the light curve, and $\sigma_i$ is the local noise (root mean square) around the pixel coordinates of the $i$-th image. For each beam, the pipeline generates a $\eta (x_p,y_p)$ map with dimensions 2800 $\times$ 2800 pixels. 

The second parameter was the S/N of the peak flux density of each light curve, which is defined as
\begin{equation}
    {\rm S/N_{peak}} = \frac{S_{i,{\rm peak}}}{\sigma_i}.
\end{equation}
Similarly, a peak map of the same dimension was generated for each beam.

If a pixel showed variability during the observation, its $\eta$ value would significantly exceed 1 as the flux density varied from the mean flux density. A transient or pulse was characterised by a high flux density compared with nearby noise, which was reflected by its high peak S/N value.

After the statistical maps were generated for each beam, pixel coordinates with value greater than 6-sigma in logarithmic space in each map were selected by the pipeline. Several additional criteria were implemented in the VASTER pipeline to filter the selected pixels:
\begin{enumerate}
  \item The pixel must be contained within the primary beam.
  \item The separation between the pixel's coordinates and the nearest catalogued source in the deep image is more than $30''$ to reduce false detections from nearby source.
  \item If (ii) is not satisfied, then
    \begin{enumerate}
        \item[(a)] the modulation index of the light curve $m=\sigma_s/\bar{S}$ is greater than 0.05, where $\sigma_s$ is the standard deviation of the light curve, and
        \item[(b)] the ratio of integrated to peak flux density is less than 1.5.
    \end{enumerate}
\end{enumerate}
Finally, the VASTER pipeline stored the coordinates of the selected local maxima from each of the statistical map as the candidates' positions. It combined the candidates into a candidate list from both maps using a $10''$ cross-matching radius. For each candidate, VASTER also generated a cropped deep image around the candidate's coordinates, the light curve $S(x_p, y_p)$, and a GIF animation showcasing all short images at the candidate's coordinates.

\subsection{Candidate classification and inspection}\label{sec:classification}
A candidate list for each field typically had several hundreds of candidates per EMU field, requiring significant time and resources for identification. We first we cross-matched all candidates' coordinates (i.e., local maxima in the statistical maps generated by VASTER) with the SIMBAD database \citep{Simbad}. We adopted a 10 arcsec cross-matching radius, which is five times the estimated astrometric uncertainty of EMU observation (${\sim}~2''$) \citep{2025PASA...42...71H}. When multiple sources were found within this radius, the closest match was selected. We inspected the light curve of those candidates. If a pulse-like structure was identified, we would generate the dynamic spectrum of the candidate.

\begin{figure*}
    \centering
    \includegraphics[width=0.95\linewidth]{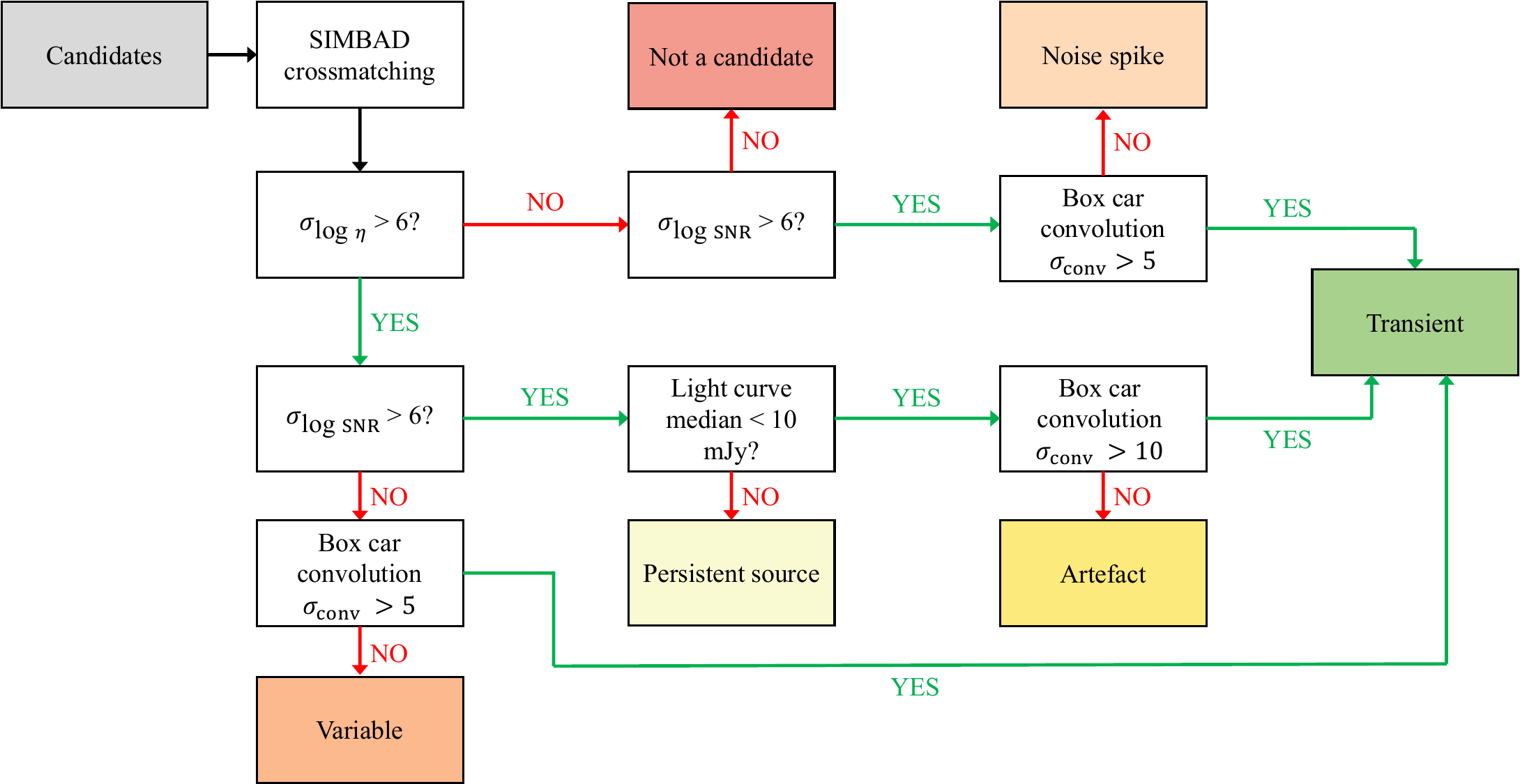}
    \caption{Flowchart of the candidate classification pipeline. olive arrows indicate a true outcome, while red arrows indicate a false outcome. Candidates that pass only one of the two statistical tests are convolved with five boxcar filters using a 5$\sigma_{\rm conv}$ threshold. Candidates that pass both tests ($\sigma_{\rm log~\eta} > 6$ and $\sigma_{\rm log~S/N} > 6$) are first filtered based on their median flux density. Those with median flux below 10 mJy are further convolved using a 10$\sigma_{\rm conv}$ threshold. Candidates classified as Transients will be inspected manually while the rest are discarded. Figure \ref{fig:lcexample} shows two example light curves for each type of candidates.}
    \label{fig:flowchart}
\end{figure*}

We then used an additional pipeline to further filter out candidates using the median flux density of the light curve and boxcar convolution. Figure \ref{fig:flowchart} shows a flow chart of the classification pipeline. The goal is to filter out persistent sources and search for pulse-like structure in the light curve. Figure \ref{fig:lcexample} shows typical examples of light curves for the different classes of candidates determined by the pipeline.

We first filtered out persistent sources using the median value of the light curve. If a candidate passed both statistical tests with its median light curve flux density greater than 10 mJy, we classified it as a "persistent sources" and would not further inspect those candidates.

We then applied five boxcar convolutions ($2^N$ time-steps, where $N=2-6$) to the light curve of the remaining candidates to search for pulse-like structure on different timescales. We used the modified z-score ($\sigma_{{\rm conv}}$) to identify peaks from the convolved light curve given by \cite{Hoaglin2013Volume1H}
\begin{equation}
    \sigma_{{\rm conv}} = \frac{0.6745(S_{{\rm conv},j}-\Hat{S}_{\rm conv})}{\rm MAD},
\end{equation}
where $S_{{\rm conv},j}$ is the $j$-th data point of the convolved light curve, $\Hat{S}_{\rm conv}$ is the median of the convolved light curve, and MAD is the median absolute deviation. This parameter served as a more robust statistic for detecting outliers (e.g., single flare event or pulses). The more common standard deviation was not chosen as it could be impacted by extreme values such as a bright pulse, and the convolved light curve is not a normally distributed function.

We adopted a conservative threshold of $\sigma_{{\rm conv}} = 5$ for candidates that passed only one statistical test. Candidates that passed only the peak map test may exhibit narrow pulses that were too weak to affect the value of $\eta$.  Candidates passing only the chi-square test might contain broad but weak pulses, leading to high variability but low peak S/N. A relatively low threshold can include more possible transients and improve the completeness of this search. In either case, if the convolved light curve exceeded the threshold, the candidate was classified as a "transient" and proceeds to manual inspection. 

For candidates passing both statistical tests, a more stringent threshold of $\sigma_{{\rm conv}} = 10$ was adopted. We noticed that the rotating side lobes of a bright source and poor astrometry are common types of artefacts that may appear as a candidate with high chi-square and peak statistics. In addition, variability due to interstellar scintillation may occur in ordinary pulsars on ${\sim}10$-second timescales. These pulsars typically have intrinsic periods shorter than 10~s and therefore appear as persistent sources in the image plane. However, scintillation can cause temporary flux increases that pass the statistical tests, even though such sources are not the type of transients we aim to detect (i.e. sources that appear and disappear relative to the noise within the observation epoch). The adoption of a higher threshold for these candidates therefore helps to filter out both instrumental artefacts and pulsar variability, increasing the reliability of the final transient sample. Candidates with pulses identified by the boxcar convolution were classified as "transients", while those that did not pass were categorised as "artefact". 

The timestamps of every transient were saved, and the candidate list was sorted based on the pulse count on the convolved light curves. All candidates categorised as "transient" underwent manual inspection. We first reviewed the candidate's light curve and GIF animation generated by VASTER to determine whether the candidate was likely to be a transient. If a candidate appeared to be a potential new transient event, we proceeded to generate a dynamic spectrum for a more detailed inspection.We applied this method to EMU observations with known LPT detections for testing, successfully identifying the pulses, and the saved timestamps provided a preliminary estimate of the period. Artificial light curves were also injected to the the pipeline and over 90\% of the injected light curves were detected by multiple boxcars (see \ref{sec:pipeline_veri} for more details).

\section{Results}

We found six transients from cross-matching all candidates with SIMBAD, three of which were also identified by the modified z-score approach. The typical offset between the transients and their corresponding SIMBAD coordinates is $2''$, which is comparable to the positional uncertainty of EMU observations. We also identified more than 30 known pulsars and persistent radio sources, such as AGN. Although pulsars are not persistent, they do not appear as transients on the 10-second timescales probed in this work. Therefore, they do not meet the transient criteria of this study and will not be discussed further.  All six transients of interest were found to be associated with stars. Their names, coordinates, radio observation Scheduling Block ID (SBID), and statistics have been listed in Table \ref{tab:all_transients}. Additional data such as peak flux density in the 10-second image, circular polarisation fraction calculated by averaging the Stokes V dynamic spectra along the time axis, full width at half maximum (FWHM) of the pulse measured from the light curve averaged at one-minute timescale, distance from Earth according to literature or Gaia parallax measurement, and the peak emission time in UTC is provided in Table \ref{tab:transients_additional details}.

\begin{table*}
    \centering
    \caption{Transients identified in this search. We list their names, coordinates of the transient from the VASTER statistical maps, SBID where the transient was found, the significance level of the weighted reduced chi-square ($\eta$) and S/N of the peak flux density (S/N$_{\rm peak}$) in log space (see Section 2.3). We also provide the classification of the object from the SIMBAD database.}
    \begin{tabular}{c|c|c|c|c|c|c}
        \hline
        Name & RA (hh:mm:ss) & Dec (dd:mm:ss) & SBID & log$(\eta)~\sigma$ & log(S/N$_{\rm peak})~\sigma$ & SIMBAD classification \\
        \hline
        Beta Centauri & 14:03:48.9 & $-$60:22:23 & 54095 & 6.00 & 0.24 & $\beta$ Cep Variable \\
        HD 105386 & 12:08:09.0 & $-$66:59:03 & 61103 & 7.23 & 0.18 & Star \\ 
        Gaia DR3 5853594572486546176 & 14:09:40.9 & $-$64:24:02 & 53310 & 19.41 & 3.36 & High PM Star \\
        HD 110244 & 12:41:18.1 & $-$58:25:57 & 61083 & 13.76 & 2.33 & Orion Variable \\
        IO Vel & 09:38:02.9 & $-$54:13:08 & 51428 & 16.13 & 5.78 & $\alpha^2$ CV variable star \\
        UCAC4 129-071513 & 12:52:22.8 & $-$64:18:41 & 55363 & 7.10 & 3.03 & Young Stellar Object Candidate \\
        \hline
    \end{tabular}
    \label{tab:all_transients}
\end{table*}

\subsection{Beta Centauri}
A pulse from Beta Centauri was identified through crossmatching its coordinates with the SIMBAD database. After accounting for proper motion, the transient is offset by $1.8''$ from the source, which is within the positional uncertainty of EMU observation. %The system is at a distance of 110.6 $\pm$ 0.5 pc from Earth \citep{2016A&A...588A..55P}.
Beta Centauri is a triple star system that consists of two main components: the closer Aa$-$Ab binary and the wider A$-$B binary. \cite{2016A&A...588A..55P} estimated the orbital period of the Aa$-$Ab binary to be 356.915 $\pm$ 0.015 days, with both stars classified as B1 III giants \citep{1988mcts.book.....H}. The A$-$B binary has an angular separation of 1.2 arcseconds, but its period remains uncertain due to a poorly constrained eccentricity ($0.5 < e < 0.9$), leading to an inferred period of 125–220 years \citep{2016A&A...588A..55P}. While Beta Centauri has been observed in infrared, ultraviolet, and X-ray wavelengths, radio emission from the system was reported only recently by \citet{vast_memes2025}. 
%it has not been detected in the radio spectrum according to the Vizier catalogue \citep{Vizier} and is not listed in the Sydney Radio Star Catalogue \citep{SRSC}.

\begin{figure*}
  \centering
  \begin{subfigure}[b]{0.48\textwidth}
    \includegraphics[width=\linewidth]{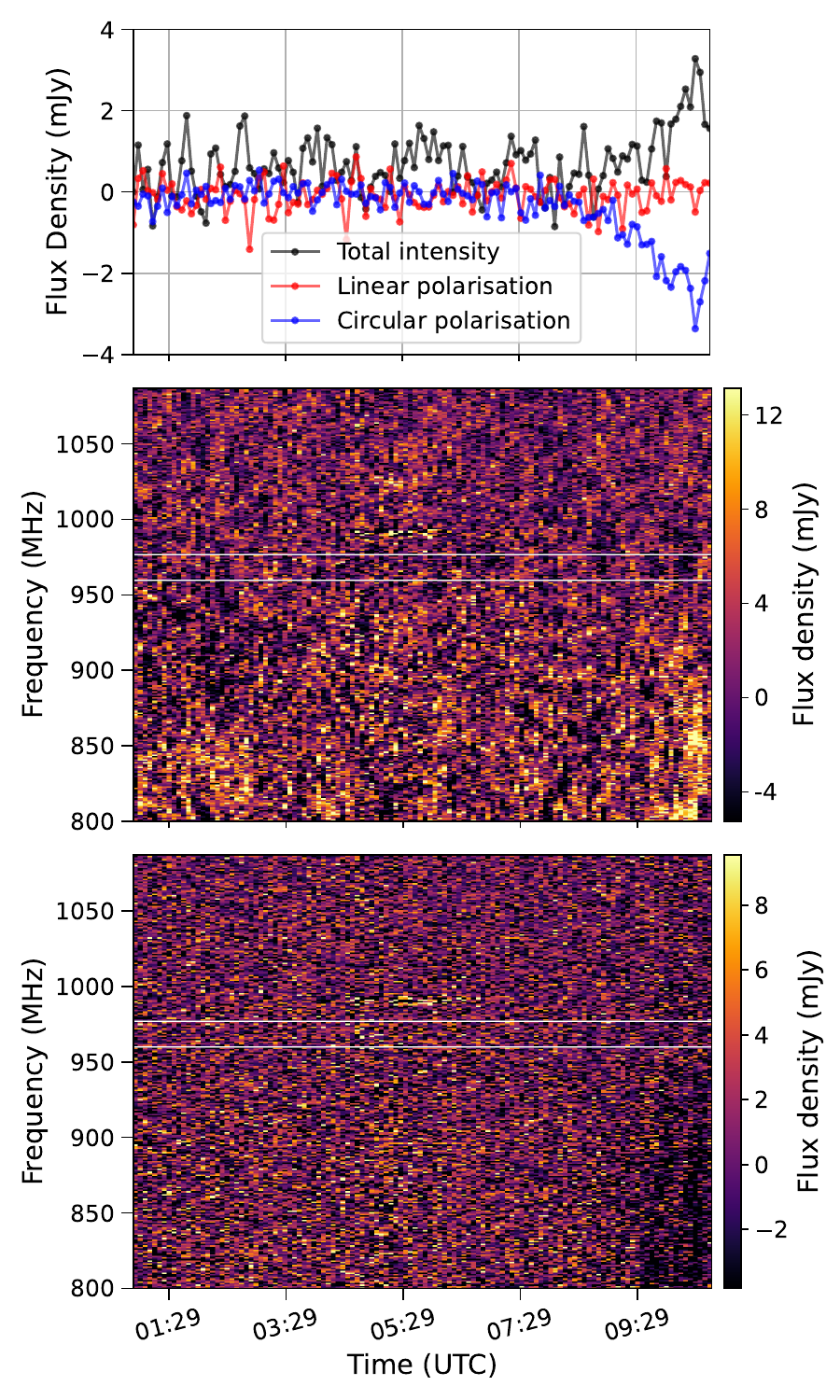}
    \caption{Beta Centauri}
    \label{fig:BetaCen}
  \end{subfigure}
  \hfill
  \begin{subfigure}[b]{0.48\textwidth}
    \includegraphics[width=\linewidth]{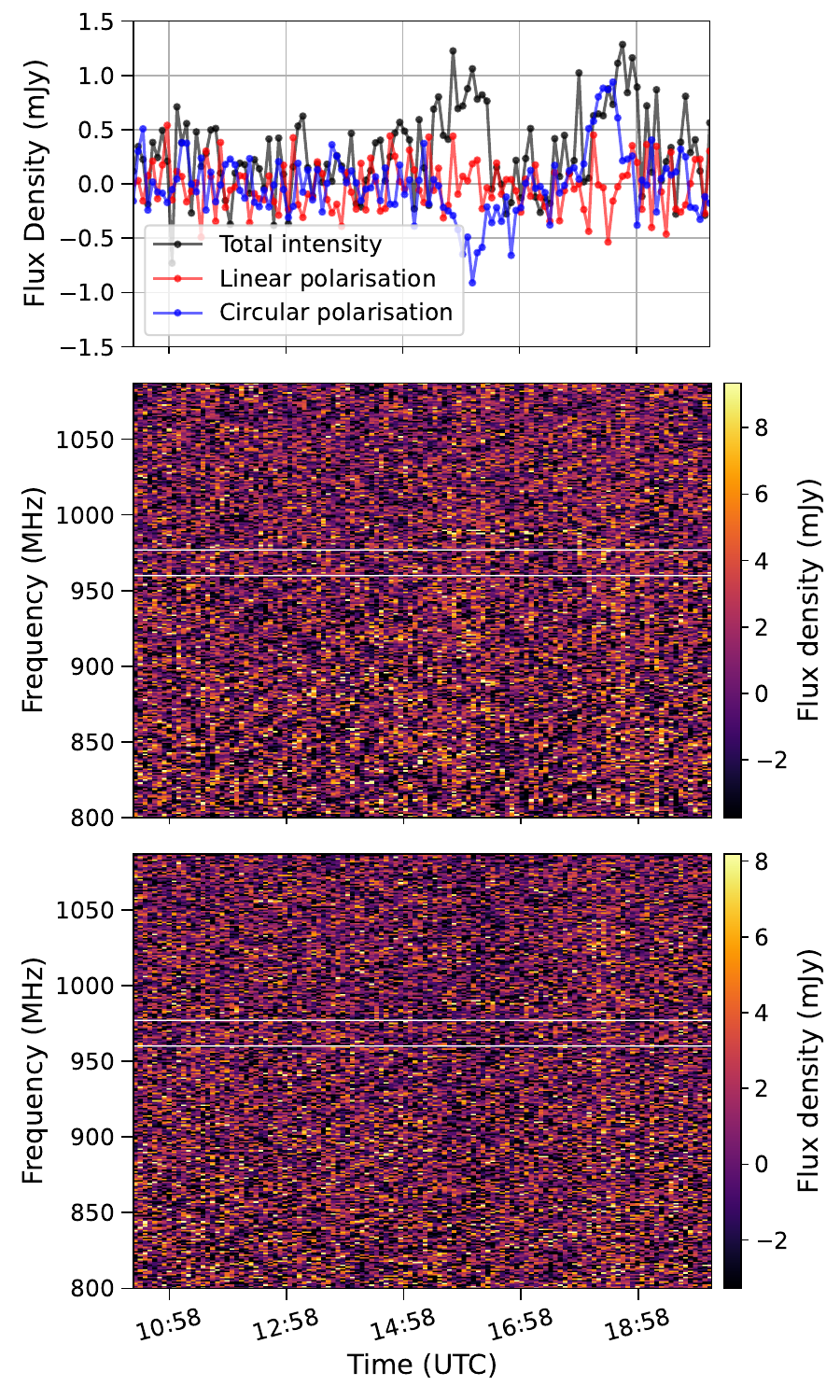}
    \caption{HD 105386}
    \label{fig:HD110244}
  \end{subfigure}
  \caption{Light curves and dynamic spectra of Beta Centauri (left) and HD 105386 (right). In both plots, the top panel shows the light curve averaged in 5-minute intervals, while the middle and bottom panels display the dynamic spectra of the Stokes I and Stokes V parameters, respectively. In panel (a), the observation ended before the full transient event could be captured. While low-level emission appears prior to the peak, its flux density is comparable to the noise level, and the Stokes V dynamic spectrum does not confirm whether these features are genuine. This transient is only detected in the lower-frequency band. HD 105386 exhibits a double-peaked structure separated by approximately three hours, with strong circular polarisation and a sign reversal between the peaks.}
  \label{fig:combined_2}
\end{figure*}

Due to the close proximity of the stars in the system, it is uncertain which one emitted the pulse. Figure \ref{fig:BetaCen} presents the light curve and dynamic spectrum of the transient, averaged over 5-minute time intervals. The transient was observed for two hours before the observation ended. Assuming the peak of the pulse occurred at the midpoint of the full transient, the pulse width is approximately four hours. This pulse was missed by the boxcar convolution due to its extended duration. It exhibits strong circular polarisation, similar to other transients detected in this study, and is only emitted in the lower band of the observed frequency range. From the light curve and Stokes I dynamic spectrum, there appears to be low-level radio emission before the transient. However, the S/N is too low, and the Stokes V dynamic spectrum cannot confirm whether these emissions are genuine.

%{\textit{Barnali: I am just inserting some text, feel free to edit it or reorganise it. Let me know if you have any questions.}\\

The highly circularly polarised radio emission from the system could be generated by centrifugal breakout (CBO), which occurs when the extent of the co-rotating magnetospheres exceeds the Kepler radius \citep[for a detailed review, see][]{townsend2005,shultz2020,owocki2020}. A subset of the magnetic hot stars, now called `Main-sequence Radio Pulse emitters' (MRPs) \citep{2021ApJ...921....9D}, can produce circularly polarised radio emission through ECME mechanism during CBO, which appear to be pulsed due to beaming effect. The secondary star of Beta Centauri (Ab) has a magnetic field strength of $230\pm 40$ G \citep{shultz2019c}, which can support a magnetosphere for CBOs to occur. The 
high circular polarisation and the pulsed nature of the observed radio emission is consistent with that expected for ECME \citep[also suggested by][]{vast_memes2025}. The observed burst could be a periodic radio pulse from the rotational timescale of the star \citep[2.885 days,][]{shultz2018}, which can be verified by conducting follow-up observations.

\subsection{HD 105386} 
The transient was identified by the widest boxcar in SB61103. It is found to be offset by $1.20''$ from the coordinates of HD 105386 in SIMBAD. This star is classified as an ApSi star, a type of chemically peculiar main-sequence star \citep{1975mcts.book.....H}. %The Gaia DR3 parallax is 2.32$\pm$0.02 mas, corresponding to a distance of $427.25^{+4.68}_{-3.84}$ pc \citep{vo:gedr3dist_main,2021AJ....161..147B}.
TESS observations revealed HD 105386 to be an eclipsing binary with a 0.99-day orbital period \citep{TESS}. We identify a single radio detection in RACS-high with an offset of only $0.16''$ from HD 105386 \citep{Vizier,RACS_high}. This positional difference is within the RACS astrometric uncertainty of $0.8''$ \citep{2021PASA...38...58H}, suggesting that the RACS-high detection is associated with HD 105386. The transient reported in the work is therefore likely to represent the second radio detection of this star.

Figure \ref{fig:HD105386} presents the light curve and dynamic spectrum of HD 105386, averaged over 5-minute time intervals. The 10-second peak flux density was only 3.6 mJy with a peak circular polarisation fraction of 86\%. The transient exhibits a double-peaked structure, but the boxcar convolution was only able to identify the first peak due to its low flux density. The peaks are separated by approximately 3 hours and a sign change can be seen in the circular polarisation.

% \begin{figure}
%     \centering
%     \includegraphics[width=0.8\linewidth]{HD105386.png}
%     \caption{Light curve and dynamic spectrum of HD 105386. The light curve in the top panel was averaged on a longer 5-minute time step to smooth out the noise level. The middle and bottom panel shows the dynamic spectrum of the Stokes I and Stokes V parameter, respectively. The transient showed a double-peak feature that is separated by roughly three hours. The pulse also showed strong circular polarisation and a sign change between the peaks.}
%     \label{fig:HD105386}
% \end{figure}

\subsection{Gaia DR3 5853594572486546176} 
\begin{figure*}
  \centering
  \begin{subfigure}[b]{0.48\textwidth}
    \includegraphics[width=\linewidth]{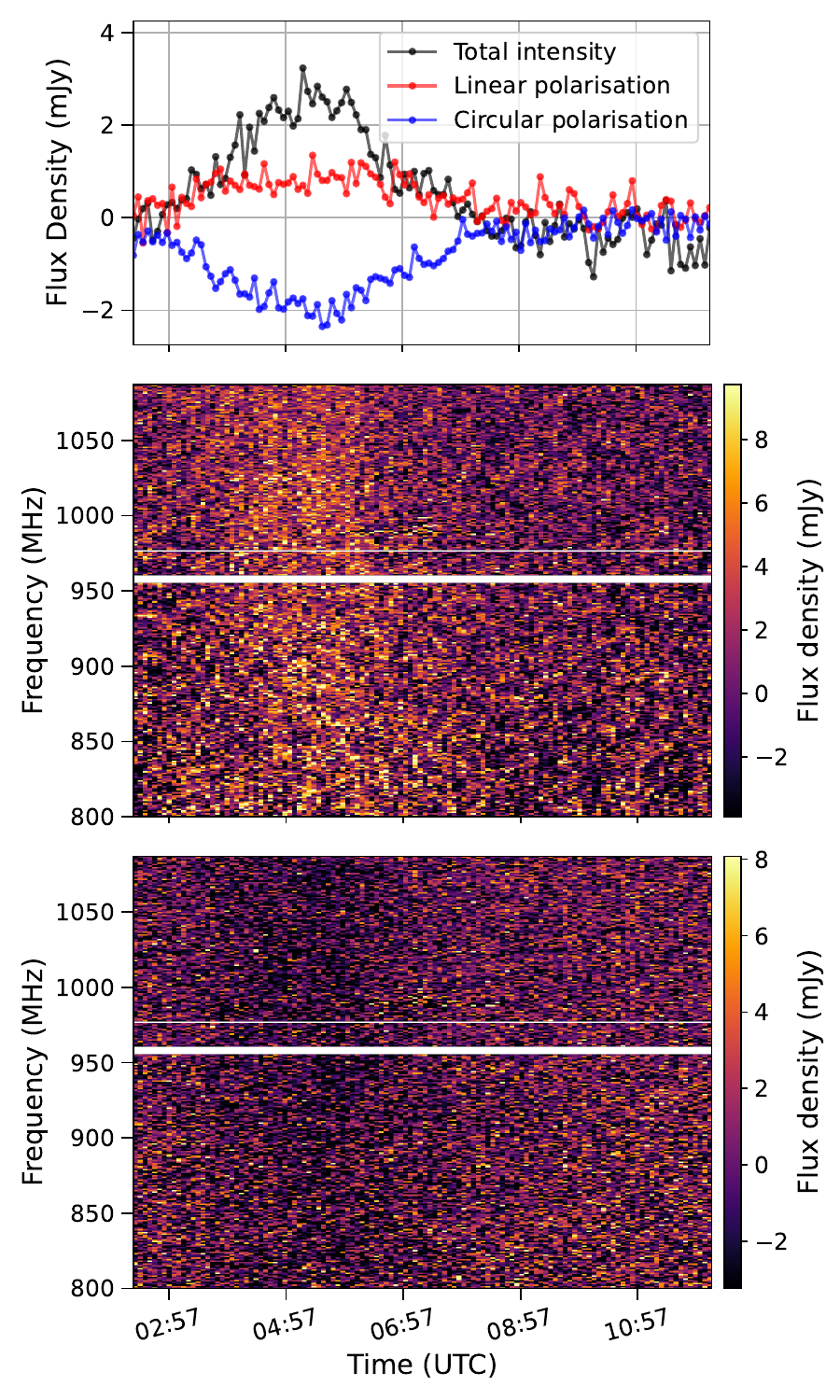}
    \caption{Gaia DR3 5853594572486546176}
    \label{fig:Gaia DR3 5853594572486546176}
  \end{subfigure}
  \hfill
  \begin{subfigure}[b]{0.48\textwidth}
    \includegraphics[width=\linewidth]{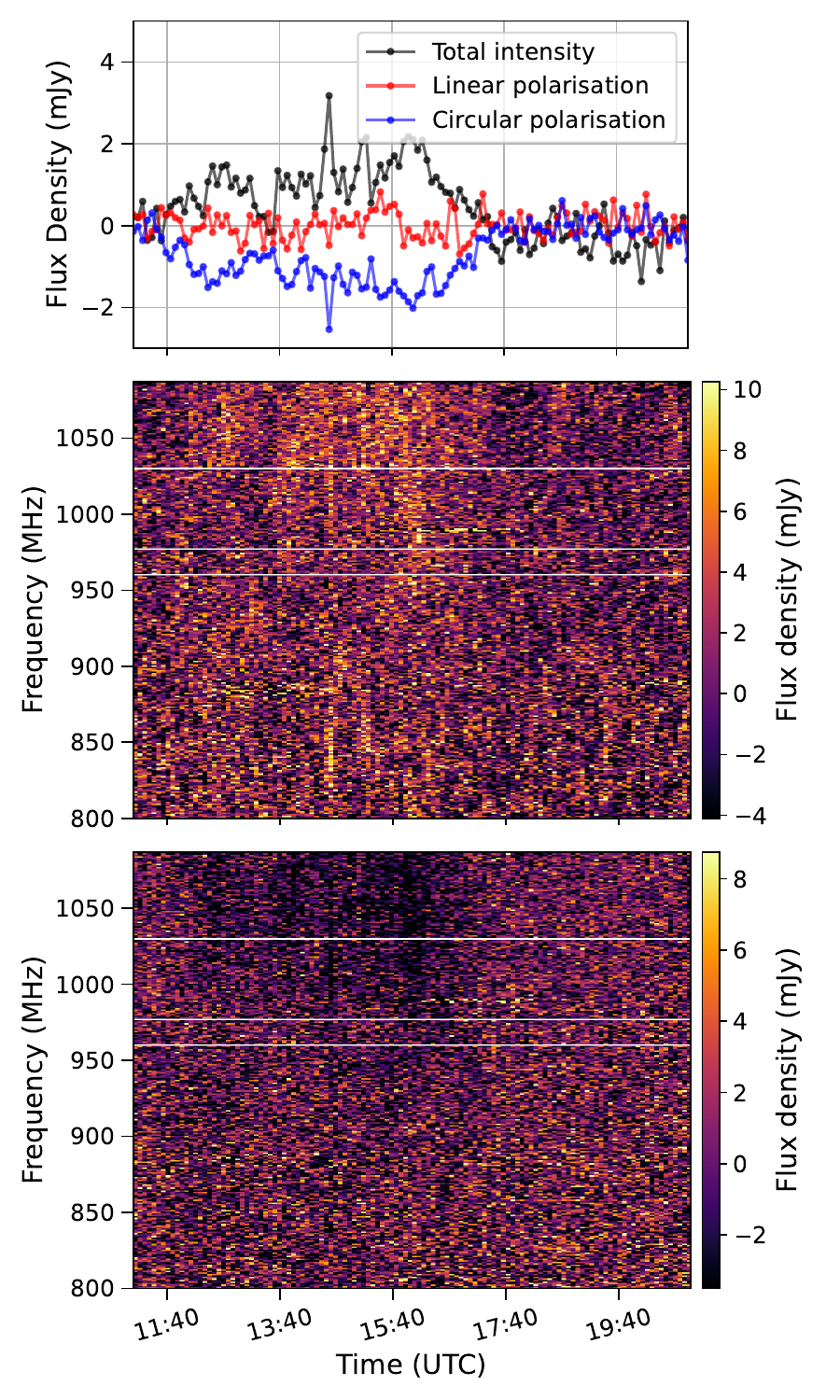}
    \caption{HD 110244}
    \label{fig:HD105386}
  \end{subfigure}
  \caption{Light curves and dynamic spectra of Gaia DR3 5853594572486546176 (left) and HD 110244 (right), with panel arrangements matching those in Figure~\ref{fig:combined_2}. The transient associated with Gaia DR3 5853594572486546176 is 50\% circularly polarised and shows a significant linear polarisation fraction, with a pulse width exceeding four hours. The transient from HD 110244 spans nearly six hours and exhibits a strong circular polarisation fraction, with emission confined to the higher-frequency bands.}
  \label{fig:combined}
\end{figure*}

Gaia DR3 5853594572486546176 was found after crossmatching the transient's coordinates with SIMBAD and was classified as a star. After accounting for proper motion, it is 1.1 arcsec away from the the coordinates in our candidate list. The source is not listed in the Sydney Radio Star Catalogue \citep{SRSC}. There were also no other published radio detections. Figure \ref{fig:Gaia DR3 5853594572486546176} presents the light curve and dynamic spectrum of the star, averaged over 5-minute time intervals. The 10-second peak flux density is 6.5 mJy and the circular polarisation fraction is 82\%. The transient is extremely broad and lasted for over 4 hours. As the maximum width of our boxcar convolution is only 64 time-steps, i.e., 640 seconds, it is not sensitive to transients that are an order of magnitude wider.

This source exhibited strong linear polarisation of about $40\%$, which is different from the other transients in this study. While stellar radio emission is typically dominated by circular polarisation, elliptically polarised pulses have also been observed from M dwarfs such as UV Ceti \citep{2019MNRAS.488..559Z} and CR Dra \citep{2021A&A...648A..13C}. According to Gaia DR3 photometry, the source has $G_{\mathrm{BP}} - G_{\mathrm{RP}} = 3.07$ and a mean $G = 14.9$. Using its measured parallax of $20.77 \pm 0.03~\mathrm{mas}$ \citep{2021AJ....161..147B}, the corresponding absolute magnitude and colour are consistent with an M5 spectral type \citep{2019AJ....157..231K}. This suggests that elliptically polarised radio emission from this source is plausible given similar behaviour observed in other M dwarfs. \citet{1991A&A...249..250M} demonstrated that elliptically polarised radio emission can arise from the electron cyclotron maser instability occurring in regions of very low electron density, which have been proposed to exist in other M dwarfs \citep[see][and references therein]{2021A&A...648A..13C}.

Some LPTs are believed to originate from white dwarf binaries with an M dwarf companion. These systems may exhibit strong linear polarisation \citep[e.g. \(L \lesssim 51\%\) in ILT~J1101$+$5521;][]{ILTJ1101}, similar to the emission observed from Gaia DR3 5853594572486546176. To explore the possibility of the transient being an LPT, we searched the Galaxy Evolution Explorer archival data \citep{2005ApJ...619L...1M} for ultraviolet counterparts. However, we did not find any within a $10''$ radius of the source. Although it is possible for a white dwarf–M dwarf binary to lack a detectable UV counterpart \citep[e.g. GLEAM-X J0704$-$37;][]{GLEAM-XJ0704‑37}, the proximity of Gaia DR3 5853594572486546176 to Earth makes it unlikely that a white dwarf companion would remain undetected in ultraviolet observations.

\subsection{HD 110244}
HD 110244 is classified as an Orion Variable by SIMBAD. It was identified during the crossmatch between our candidate list and SIMBAD. The transient is 1.7 arcsec away from the coordinates of the star after accounting for proper motion. %The Gaia DR3 parallax is 9.39$\pm$0.08 mas, corresponding to a distance of $106.68^{+0.98}_{-1.03}$ pc \citep{vo:gedr3dist_main,2021AJ....161..147B}. 
This source has a spectral type of G7 \citep{2006A&A...460..695T} with a peak flux density of 6.6 mJy in the 10-second snapshot. There was also one radio detection previously from ASKAP according to \cite{SRSC}.

Figure \ref{fig:HD110244} shows the light curve and dynamic spectrum of HD 110244, averaged over 5-minute time steps. Similar to Gaia DR3 5853594572486546176, it has a broad pulse width, lasting for nearly six hours. Therefore, it was also not identified by the boxcar convolution. The transient was highly circularly polarised and was only detected in the higher frequency part of the observing band (between 950 MHz and 1088 MHz).

\subsection{IO Vel} 
IO Vel (HD 83625) was identified in SB51428 using the modified z-score. It is an $\alpha^2$ Canum Venaticorum variable star, belonging to a broader class of Ap and Bp stars. \cite{1975mcts.book.....H} classified the star as an ApSi star, suggesting an abnormal abundance of silicon. \cite{1978A&AS...34..445R} found a period of $1.080\pm0.004$ days. Observations and analysis by \citep{2025arXiv250507195D} confirmed that this star is an MRP. IO Vel has also been detected in the infrared spectrum by Two Micron All-Sky Survey (2MASS) \citep{2MASS} and in the UV spectrum by the Celescope Catalog of Ultraviolet Stellar Observations \citep{CEL} and the Catalogue of stellar ultraviolet fluxes (TD1) \citep{TD1}. %The Gaia DR3 parallax is 5.78$\pm$0.05 mas, placing it $172.94^{+1.32}_{-1.49}$ pc away from Earth \citep{vo:gedr3dist_main,2021AJ....161..147B}. 
\cite{FORS1} estimated the star's magnetic field strength to be $1245\pm77$ G. This star has been detected as a radio star in the ASKAP VAST and RACS surveys \citep{SRSC} and by \cite{vast_memes2025}.

\begin{figure}
    \centering
    \includegraphics[width=1.0\linewidth]{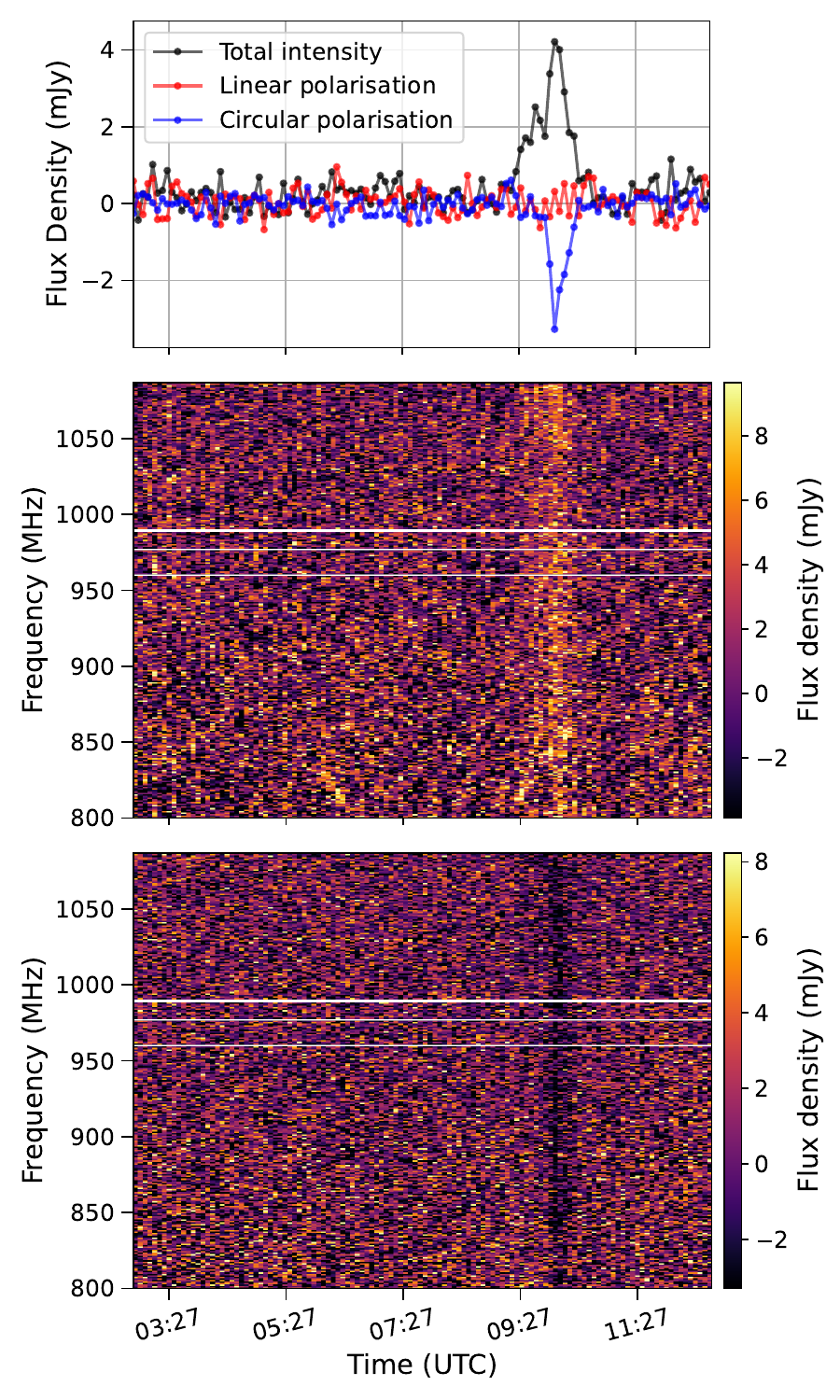}
    \caption{Light curve and dynamic spectrum of IO Vel. The top panel shows the light curve averaged in 2-minute time steps to improve visibility of the pulse, with total intensity in black, linear polarisation in red, and circular polarisation in blue. The middle and bottom panel shows the dynamic spectrum of the Stokes I and Stokes V parameter, respectively. The pulse width is around 100 minutes and the circular polarisation is 77\%.}
    \label{fig:HD83625}
\end{figure}

Figure \ref{fig:HD83625} shows the light curve and dynamic spectrum of the transient after averaging in 5-minute time intervals. The 10-second peak flux density was 7.3 mJy with a circular polarisation fraction of 73\%. The pulse duration is roughly 100 minutes.

\subsection{UCAC4 129-071513} 
UCAC4 129-071513 is a star found in SB55363 identified using the modified z-score. %The Gaia DR3 parallax is 5.78$\pm$0.05 mas, corresponding to a distance of $101.51^{+0.20}_{-0.20}$ pc \citep{vo:gedr3dist_main,2021AJ....161..147B}. 
This star has also been detected in the infrared spectrum by 2MASS \citep{2MASS} and Wide-field Infrared Survey Explorer (WISE) \citep{WISE}. Photometric data from \cite{UCAC4} and \cite{2MASS} suggest that UCAC4 129-071513 is a late K-type to M-type main-sequence star \citep{colour-magnitude-diagram}. It has been identified as a radio star in the ASKAP VAST and RACS surveys \citep{SRSC}.

\begin{figure*}
  \centering
  \begin{subfigure}[b]{0.48\textwidth}
    \includegraphics[width=\linewidth]{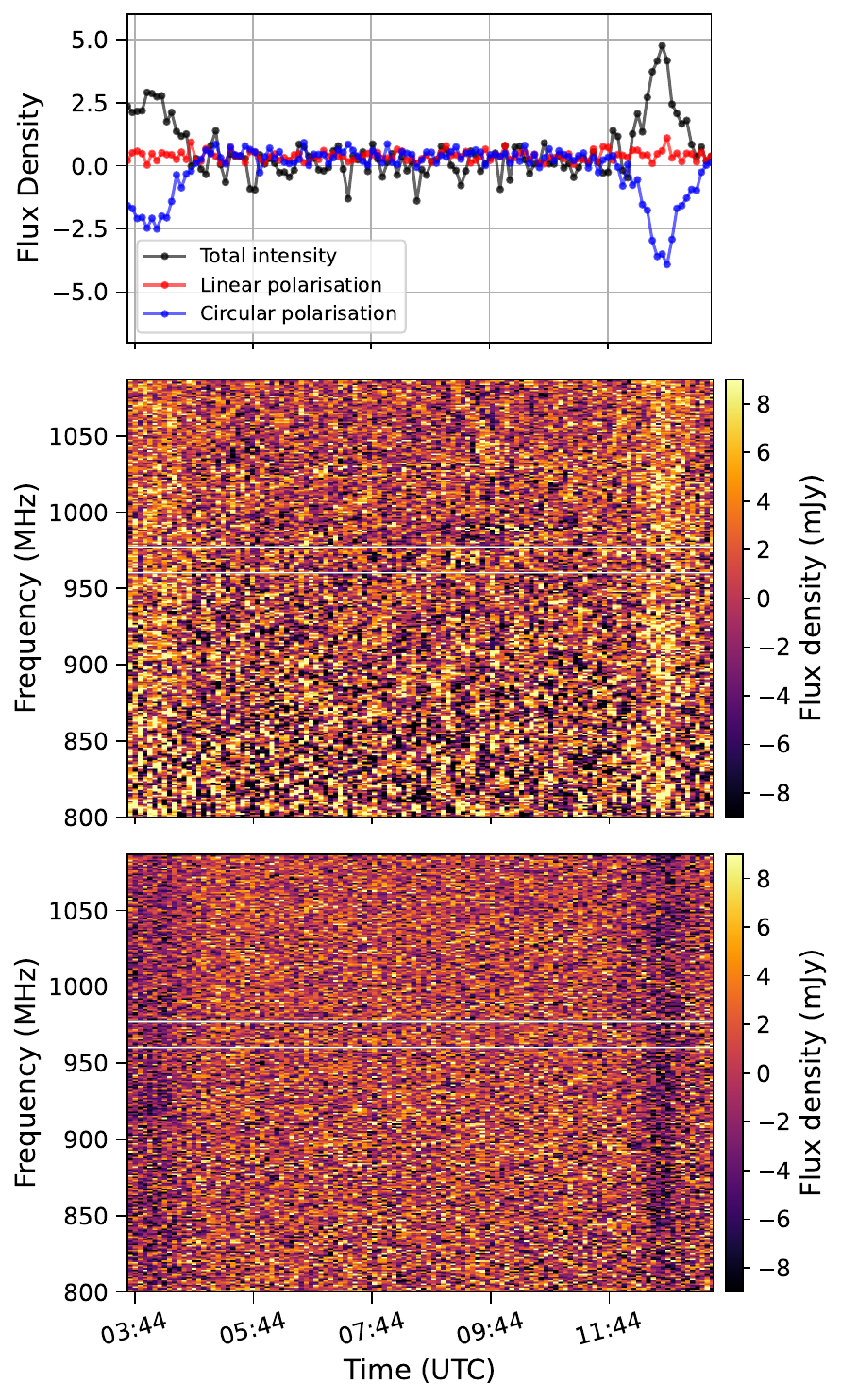}
    \caption{SB55363}
    \label{fig:SB55363}
  \end{subfigure}
  \hfill
  \begin{subfigure}[b]{0.48\textwidth}
    \includegraphics[width=\linewidth]{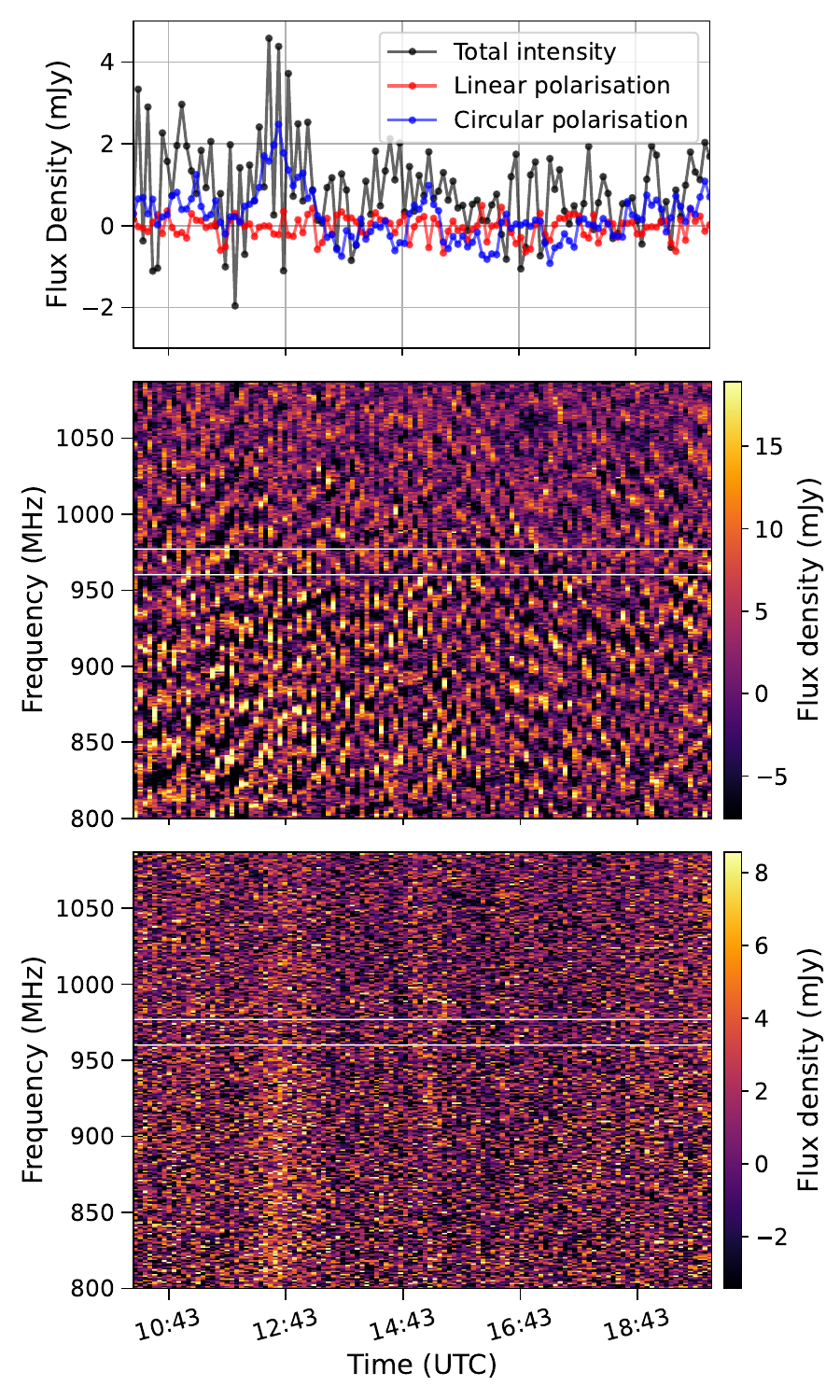}
    \caption{SB62225}
    \label{fig:SB62225}
  \end{subfigure}
  \caption{Light curve and dynamic spectrum of UCAC4 129-071513, averaged in 5-minute time steps. The left column shows the observation from SB55363, while the right shows the observation from SB62225. The top panels display the light curve and polarisation of the transient, while the middle and bottom panels show the Stokes I and Stokes V dynamic spectra, respectively. The source was identified as a candidate in SB55363 but not in SB62225, likely due to side-lobe artefacts from nearby sources.}
  \label{fig:UCAC_combined}
\end{figure*}

The source was affected by the sidelobes of nearby sources and the data had to be further modelled and subtracted using \texttt{DStools}\footnote{\url{https://github.com/askap-vast/dstools}} \citep{Pritchard2025a}. Figure \ref{fig:SB55363} presents the light curve and dynamic spectrum of the transient from SB55363, averaged over 5-minute time intervals. The transient has a peak flux density of 7.8 mJy and a circular polarisation fraction of 86\%. The pulse duration is approximately 90 minutes. A second pulse is also visible at the beginning of the observation. The coordinates of UCAC4 129-071513 were also covered by another EMU field processed in this study, SB62225, but it did not appear in the candidate list. Upon inspecting the dynamic spectrum, we identified a pulse in the Stokes V parameter. This pulse was also likely obscured by the sidelobes of nearby sources in the Stokes I parameter, as seen in the light curve in Figure \ref{fig:SB62225}. 

To estimate the source's period, we conducted Monte Carlo sampling on the transient's time of arrival, using detections from the ASKAP VAST, RACS, and EMU surveys. Given that the pulse width exceeds the integration times of VAST (12 minutes) and RACS (15 minutes), we assumed an uncertainty of 20 minutes in the time of arrivals. Despite this approach, we were unable to determine a period that consistently accounts for all detected pulses.

\section{Discussion}
This work is the first large-scale search using ASKAP data in the image plane at 10-second-timescale resolution. We note that at least one of the detected sources has not been previously detected at the radio wavelengths. Previous ASKAP radio star searches have primarily used VAST and RACS data \citep[e.g.,][]{VAST,2021MNRAS.502.5438P}, which have observation durations of only 12 and 15 minutes, respectively. Such short observation windows may have missed transients entirely or failed to capture their full duration, limiting the study of their variability. Longer observation windows not only allow for the detection of broader pulses but also improve sensitivity by reducing the noise level in deep images. Future searches targeting radio stars or MRPs can leverage ASKAP EMU and Widefield ASKAP L-band Legacy All-sky Blind surveY \citep[WALLABY;][]{2020Ap&SS.365..118K} data products. WALLABY observes using a central frequency of 1367.5 MHz for 8 hours per pointing.

Three out of the six transients discovered in this search were not identified by the boxcar convolution but instead through cross-matching with the SIMBAD database and manual inspection of the dynamic spectrum. There are two key reasons for this. Firstly, the boxcar configurations were not optimised for detecting hour-long transients. This search aimed to identify more isolated LPTs. The pulse widths of hour-long LPTs range from tens of seconds \citep{ASKAPJ1935} to several hundred seconds \citep{ASKAPJ1839}. We tested the boxcar convolution and modified z-score statistics using known LPT sources, successfully identifying their pulses, so the pipeline is effective but only up to a certain pulse width. Secondly, the transients exhibit relatively weak peak flux densities compared to the noise level. As shown in Table \ref{tab:all_transients}, none of the detected transients have log(S/N$_{\rm peak})~\sigma$ values exceeding the detection threshold. This may explain why their modified z-scores also fall below the cut-off. This issue is worse for wider pulses that span a significant fraction of the observation (e.g., Gaia DR3 5853594572486546176 and HD 110244), as the transient itself increases the median flux level of the light curve, further reducing its detectability.

\subsection{Detectability of LPTs in this work}

\subsubsection{Pipeline verification} \label{sec:pipeline_veri}
This search targeted the Galactic plane because most isolated LPTs discovered to date lie at low Galactic latitudes. Yet no new LPTs were detected. Before attributing this result to the intrinsic properties of LPTs, we first verified that our pipeline could in fact recover these sources under conditions similar to those of this search.

We processed four additional ASKAP observations that included two previously known LPTs: ASKAP J1935$+$2148 and ASKAP J1839$-$0756. The former exhibits a period of 54 minutes and pulse widths ranging from 10 to 50 seconds \citep{ASKAPJ1935}, placing it near the lower limit of the timescales probed in this study. The latter LPT, with a period of 6.45 hours and pulse widths of up to 710 seconds, has the longest period and broadest pulse known to date. \citep{ASKAPJ1839}. Each observation lasted over eight hours, used the same 36-beam footprint as the EMU survey, and was positioned at low Galactic latitude. In all cases, VASTER successfully detected the LPTs, with both statistical tests exceeding the $6\sigma$ threshold. The candidate classification pipeline correctly identified the pulses and recovered the period of the LPTs from the time stamps. Additionally, the detection of the stellar flare from IO Vel -- with a pulse width of several thousand seconds -- demonstrated that our pipeline is sensitive to the timescale of interest (10–1000 seconds). Therefore, LPTs of similar pulse width at low Galactic latitudes should have been detectable in this search. 

To assess the detection efficiency of the boxcar convolution pipeline, particularly because broader pulses have been missed in our search, we carried out an injection–recovery test using artificial light curves with a range of pulse widths and amplitudes. We assumed that all injected light curves represent transients that would already have passed at least one of the statistical tests in the VASTER pipeline, and thus be treated as candidates.

We first constructed a set of 10 noise-only light curves by selecting non-candidate pixels from the searched fields. These curves show no intrinsic variability or transient events, ensuring that the background noise level is representative of genuine candidates. Each simulated light curve was created by adding a single Gaussian pulse to a randomly selected noise curve. The pulse amplitude was drawn uniformly between 3 mJy and 23 mJy, and the pulse width was drawn uniformly between 50 and 2050 seconds. The pulse centre was placed at a random position within the light curve, allowing for the possibility that some pulses are only partially contained within the observation window.

To systematically test the detection performance, we divided the amplitude–width parameter space into a $10\times10$ grid (10 bins in amplitude and 10 bins in width), resulting in 100 distinct parameter bins. For each bin, we generated and injected 1000 artificial light curves into the boxcar convolution pipeline. A pulse was considered recovered if the timestamp reported by the pipeline lay within one boxcar width of the injected pulse’s true timestamp. We evaluated the recovery rate for all boxcars (40 seconds to 640 seconds) and both z-score thresholds: $\sigma_{{\rm conv}} = 5$ and $\sigma_{{\rm conv}} = 10$. Table \ref{tab:recoveryrate} summarises the overall recovery percentages for all injected light curves, grouped by boxcar width and detection threshold. The recovery rate within each amplitude–width bin for each boxcar width and threshold is shown as a series of heat map in Figure \ref{fig:recovery_heatmap}.

\begin{table}
    \centering
    \begin{tabular}{|c|c|c|c|c|c|}
        \hline
        Boxcar width (s) & 40 & 80 & 160 & 320 & 640 \\
        \hline
        $\sigma_{{\rm conv}} = 5$ & 85\% & 87\% & 91\% & 95\% & 97\% \\
        $\sigma_{{\rm conv}} = 10$ & 65\% & 73\% & 83\% & 91\% & 94\% \\
        \hline
    \end{tabular}
    \caption{Average pulse recovery rate using various boxcar width and z-score threshold. The detailed recovery rate within each amplitude–width bin is provided in figure \ref{fig:recovery_heatmap}.}
    \label{tab:recoveryrate}
\end{table}

We find that the recovery rate generally improves when wider boxcars are used, likely because broader boxcars average out noise more effectively and improve the statistical significance of the pulse. In most cases, over 90\% of pulses are detected by at least two different boxcar widths. The $6\sigma$ detection threshold adopted in the VASTER pipeline corresponds to a sensitivity of 14.6 mJy (see Section~\ref{subsec:sensitivity} for details). From Figure \ref{fig:recovery_heatmap}, we see that the pipeline recovers more than 80\% of pulses at this flux density, regardless of pulse width or boxcar width. Even for significantly lower flux densities (4~mJy, which is only $1.6\sigma$ above the noise) and broader pulse widths (${>}1000$~s), the 64-time-step boxcar still recovers over 90\% of pulses in most scenarios. We therefore conclude that the pipeline can detect radio pulses across a wide range of parameter space with high successful rate. However, it should be noted that the pipeline may not be fully sensitive to all transients, particularly those with low flux densities or short durations, as demonstrated by the missed detection of the second peak of HD 105386.

\subsubsection{Luminosity limit from a single 10-second image} \label{sec:L_limit_image}
Given the absence of new LPTs in our search, we attempt to place constraints on the luminosity of sources that would have been detectable by our searches. We estimated the minimum radio luminosity required for a source to be observed under our survey specifications. We approached this from both the image-plane analysis and the time-series analysis perspectives.

In this approach, we assumed a transient would be detectable if its peak flux density in a 10-second image exceeded the $6\sigma$ detection threshold, which corresponds to 14.6~mJy  (see Section~\ref{subsec:sensitivity} for details). For comparison with the historic literature, we consider that LPTs have the same beam geometry as canonical pulsars. In this case, the luminosity $L$ of the source is given by
\begin{equation}
    \label{eq:luminosity}
    L = 2\pi d^2(1 - \cos\rho)\, S_{\rm peak}(f_0) \, \frac{f_0^{-\alpha}}{\alpha+1} \left(f_2^{\alpha+1} - f_1^{\alpha+1} \right),
\end{equation}
where $d$ is the distance to the source, $\rho$ is the beam opening angle, $S_{\rm peak}(f_0)$ is the peak flux density at reference frequency $f_0$, $[f_1, f_2]$ is the observing bandwidth, and $\alpha$ is the spectral index \citep{Handbook_of_Pulsar}. For an EMU field, we adopted $f_0 = 943.5~\mathrm{MHz}$, $f_1 = 800~\mathrm{MHz}$, and $f_2 = 1088~\mathrm{MHz}$. Despite the wide range of reported spectral indices for LPTs ($-3.17 < \alpha < +0.4$) \citep{GPMJ1839-10,ASKAPJ1935}, the spectral index-dependent term in Equation~\ref{eq:luminosity} varies by less than 5\%. The beam opening angle $\rho$ is less well constrained for LPTs and may depend on their periods and period derivatives \citep{2022MNRAS.514L..41E}. 

\begin{figure}
    \centering
    \includegraphics[width=1.0\linewidth]{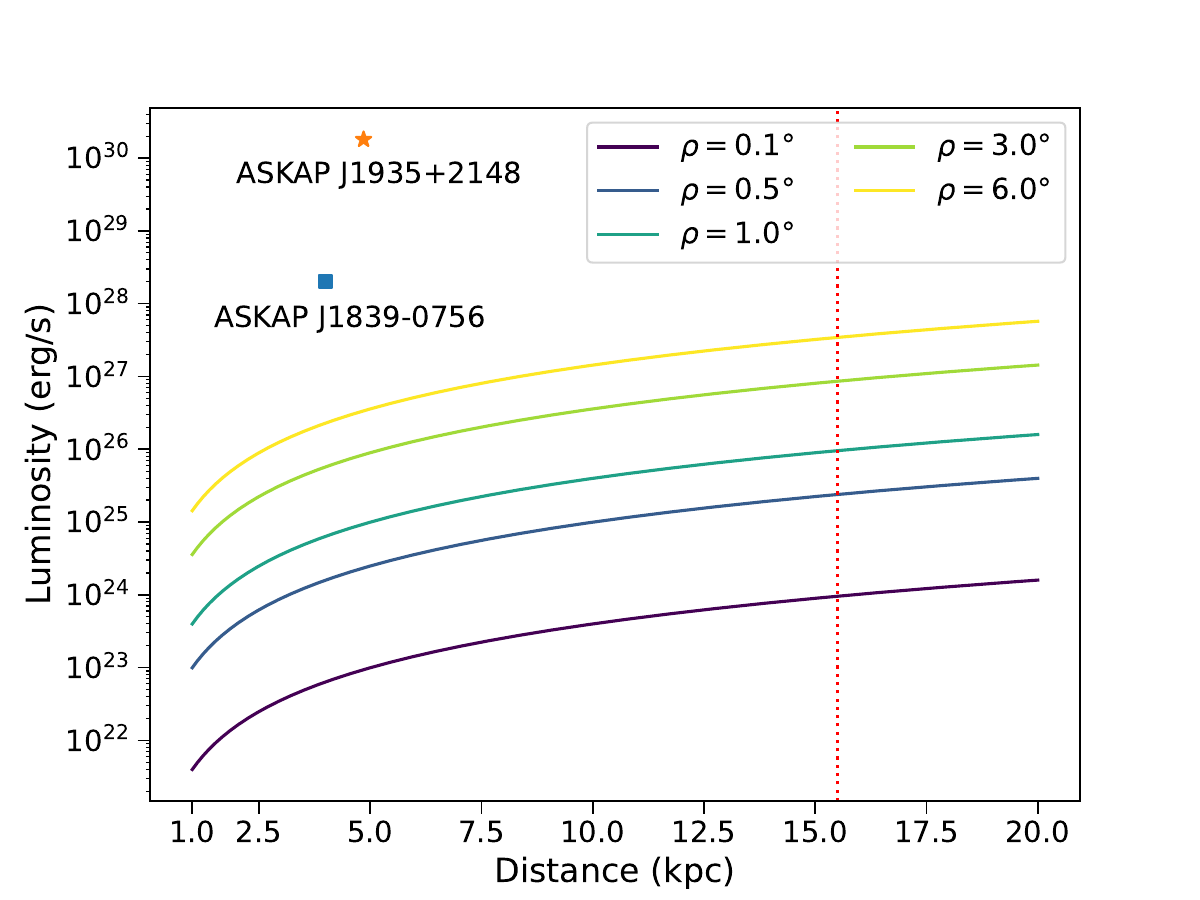}
    \caption{Minimum luminosity required for a source to be detectable in this work as a function of distance, shown for various beam opening angles $\rho$. The vertical dashed line at 15.5~kpc marks the distance from Earth to the edge of the Galaxy in the direction of Galactic longitude $\ell = 300^\circ$. Also shown are the luminosities of ASKAP J1935$+$2148 \citep{ASKAPJ1935} and ASKAP J1839$-$0756 \citep{ASKAPJ1839}, two LPTs used to validate the pipeline’s capability to detect such sources.}
    \label{fig:image_plane_luminosity}
\end{figure}

Figure~\ref{fig:image_plane_luminosity} shows the luminosity as a function of distance for various beam opening angles, ranging from $\rho = 0.1\degree$ to $\rho = 6\degree$. The lower limit corresponds to the expected beam opening angle of a pulsar with a period of one hour, based on the relation
\begin{equation}
    \rho = 6\degree \times P^{-1/2},
\end{equation}
where $P$ is the period in seconds \citep{Handbook_of_Pulsar}. We note that this empirical relation, derived from the pulsar geometry, may not be directly applicable to LPTs. In fact, the actual beam opening angles of LPTs could be an order of magnitude wider \citep[see, e.g.,][]{ASKAPJ1839}. The upper limit of $\rho = 6^\circ$ represents a generic beam width adopted for estimating LPT luminosities independent of their spin period \citep[e.g.,][]{GLEAM-XJ1627,ASKAPJ1832-0911}. Since most EMU fields in this study are clustered around Galactic longitude $\ell \approx 300\degree$, we estimated the distance from Earth to the edge of the Galaxy in this direction to be approximately $15.5~\rm{kpc}$. Using small angle approximation, we further simplified the luminosity limit to
\begin{equation}
    \label{eq:simplified_luminosity}
    L_{\rm min} = 9.4\times 10^{25} \left( \frac{d}{15.5~\rm kpc} \right)^2 \left( \frac{\rho}{1\degree} \right)^2 ~\rm erg/s
\end{equation}

We note that this calculation does not account for dispersion, which can smear a pulse over multiple images and reduce its flux density in any single 10-second image. To quantify the effect of dispersion, we consider the simplest case in which a pulse is spread over two consecutive images, corresponding to a dispersion-induced delay of 10~s. The relationship between time delay ($\Delta t$) and DM is given by \cite{Handbook_of_Pulsar}
\begin{equation}
    \Delta t = 4.15~{\rm ms} \times \left[ \left( \frac{f_1}{\rm GHz} \right)^{-2} - \left( \frac{f_2}{\rm GHz} \right)^{-2} \right] \times \left( \frac{\rm DM}{\rm cm^{-3}~pc} \right),
\end{equation}
where $[f_1, f_2]$ is the observing bandwidth. In the EMU band, a pulse with ${\rm DM} \gtrsim 3360~\rm cm^{-3}~pc$ will be smeared across two images, effectively halving its flux density in each. The Galactic electron density models NE2001 \citep{NE2001} and YMW16 \citep{YMW16} estimate that, in the direction of $\ell \approx 300\degree$ and $b=0\degree$, the total Galactic DM at a distance of 15.5~kpc is well below $1000~\rm cm^{-3}~pc$. Therefore, the reduction in flux density due to dispersion is negligible for the purposes of our luminosity limit calculation.

\subsubsection{Luminosity limit from convolving light curve}
The single 10-second image approach does not account for pulse width. By convolving the light curve with wider boxcar functions, the noise is smoothed out, allowing weaker, broader pulses to be detected. Using the typical noise level in the light curve and the more stringent $10\sigma$ detection threshold adopted in section~\ref{sec:classification}, we estimated a sensitivity of 15~mJy for a 10-second pulse. This sensitivity ($S_{\rm lim}$) improves as the noise decreases with timescale ($T$), which can be expressed as 
\begin{equation}
    \label{eq:sensitivitylimit}
    S_{\rm lim} = 15~{\rm mJy} \times \sqrt{\frac{10~\rm s}{T}}.
\end{equation}
We replace the peak flux density $S_{\rm peak}(f_0)$ in Equation~\ref{eq:luminosity} with equation \ref{eq:sensitivitylimit} to estimate the minimum luminosity detectable as a function of pulse width. Using the same distance and small angle approximation in Section \ref{sec:L_limit_image}, this limit can be rewritten as
\begin{equation}
    L_{\rm min} = 9.7\times 10^{25} \left( \frac{d}{15.5~\rm kpc} \right)^2 \left( \frac{T}{10~\rm s}\right)^{-1/2} \left( \frac{\rho}{1\degree} \right)^2 ~\rm erg/s
\end{equation}

\begin{figure*}
    \centering
    \includegraphics[width=1.0\linewidth]{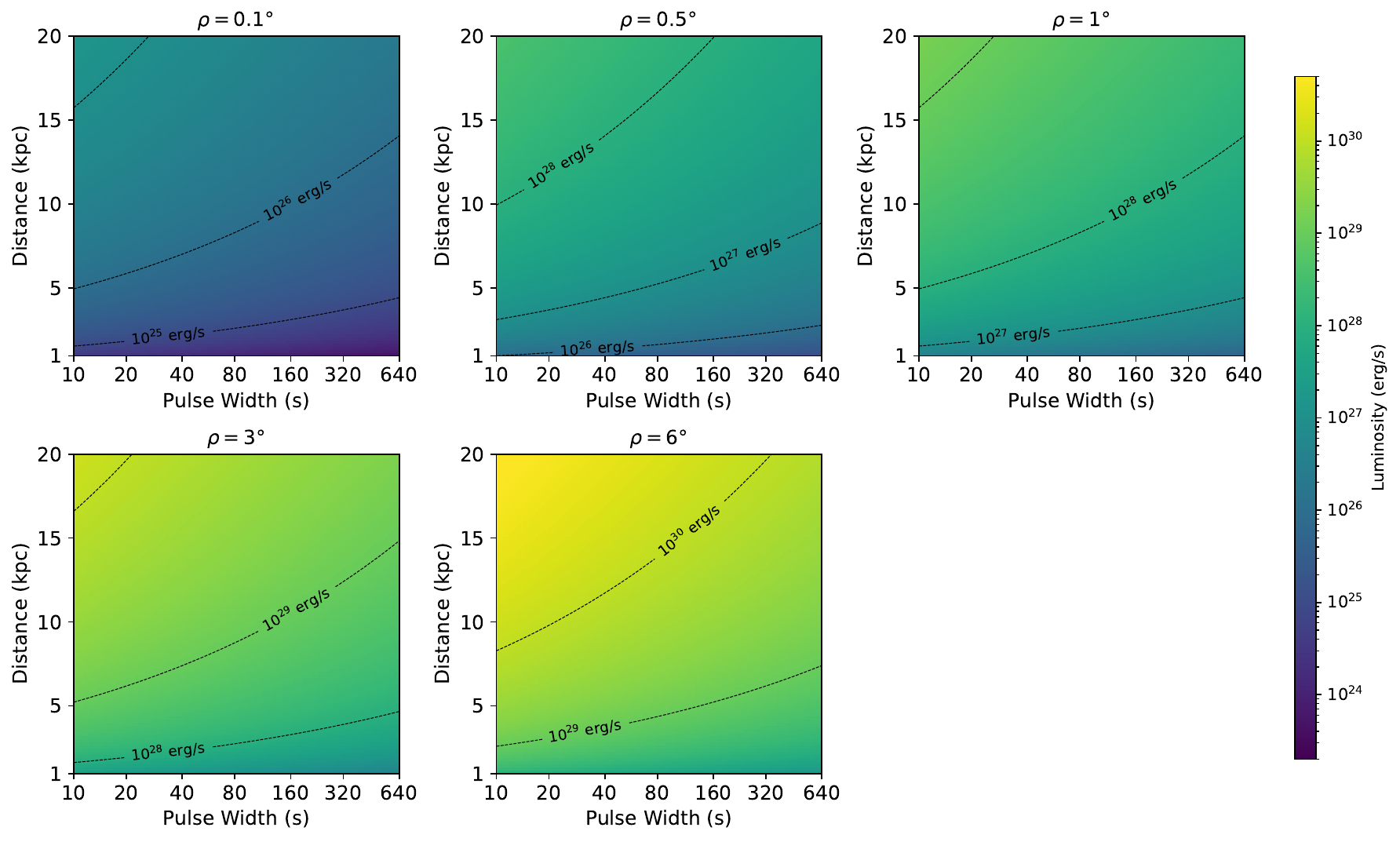}
    \caption{Heat maps showing the minimum luminosity required for detection. This is shown as a function of the pulse width in logarithmic scale (x-axis) and distance in linear scale (y-axis). Each panel represents a different beam opening angle ranging from $\rho=0.1\degree$ to 6$\degree$, increasing from top-left to bottom-right. Black contour lines in each plot mark constant luminosity levels for better visualisation.}
    \label{fig:colour_map}
\end{figure*}

The luminosity limit for various distances, pulse width, and beam opening angles are shown in Figure \ref{fig:colour_map}. The beam opening angle, $\rho$, which is tied to the emission mechanism and geometry of LPTs, has a strong influence to the luminosity of the source. Additionally, $\rho$ is linked to the pulse width via the source's period (minutes to hours) and duty cycle \citep[less than 1\% to 20\%,][]{GPMJ1839-10,ASKAPJ1935}. However, due to the limited known population of LPTs, there is currently no robust statistical model describing these parameters. A comprehensive simulation of the evolution of the LPT population, incorporating their period, age, and magnetic field, lies beyond the scope of this work and warrants a separate study. As a result, while we are able to place constraints on the luminosity of detectable sources, we are unable to constrain other intrinsic physical parameters in this study.

% \begin{figure}
%     \centering
%     \includegraphics[width=1.0\linewidth]{Figures/Fluence_completeness.png}
%     \caption{Peak flux density of a pulse as a function of its width. The blue solid diagonal lines indicate constant fluence levels, while the red dotted line marks the 10$\sigma$ sensitivity threshold from the boxcar convolution applied to the light curve. The maximum boxcar width used in the search is 640 seconds, indicated by the black dot-dashed line. The orange dashed line shows the fluence completeness limit of this work at 1.32 Jy s.}
%     \label{fig:fluence}
% \end{figure}

\subsubsection{Lack of new LPT detections}
The LPTs used to validate the pipeline -- ASKAP J1935$+$2148 and ASKAP J1839$-$0756 -- have luminosities a few orders of magnitude above our derived detection limit. This accounts for their successful detections in both the image-plane and time-series analyses. Our luminosity limit also falls below that of other known LPTs, whether isolated or in binary systems. For example, GPM J1839$-$10 has a luminosity of $L \approx 10^{28}~\rm erg\,s^{-1}$, while GLEAM-X J0704$-$37 has $L \sim 10^{26}~\rm erg\,s^{-1}$, both are at a much closer distance than the limit we set (15.5 kpc). Therefore, LPTs with distances and luminosities comparable to the current population would be detectable in our search. The absence of new detections therefore likely reflects the intrinsic rarity of LPTs, activity states of LPTs, as well as potential biases in field selection, rather than limitations of the detection pipeline.

% DD attempted rewrite: Replace "The current population...within the thin disk of the galaxy." with the following
The population of LPTs is currently thought to arise from two progenitor classes: white dwarf binary systems and isolated compact objects. GLEAM-X J0704$-$37, ILT J1101+5521, and CHIME/ILT J1634$+$44 have known optical counterparts and are confirmed white dwarf binary systems \citep{GLEAM-XJ0704‑37,2025A&A...695L...8R, ILTJ1634+44, CHIMEJ1634+44}. The remaining sample of LPTs do not have optical/infrared counterparts and appear to be isolated. This is because their low galactic latitudes (see Figure~\ref{fig:EMU_field}) limit the detectability of any counterpart due to dust extinction or crowded stellar fields.
However, all LPTs discovered to-date lie comfortably within the thick disk of the galaxy, and most are within the thin disk \citep{ASKAPJ1755}.
% The current population of LPTs is thought to consist of two main classes of progenitors: white dwarf binary systems and possibly isolated compact objects. Given the small sample size, white dwarf binaries such as GLEAM-X J0704$-$37 and CHIME/ILT J1634$+$44 appear to lie outside the thin Galactic disc, making their optical/infrared counterparts less likely to be obscured by dust or crowded stellar fields. % Because the fields targeted in this study are concentrated at low Galactic latitudes, such systems are unlikely to be detected here.
% In contrast, LPTs without known electromagnetic counterparts are found at low Galactic latitudes (see Figure~\ref{fig:EMU_field}) and, those with distances tend to lie within the thin disk of the Galaxy \citep{ASKAPJ1755}.
The distance from the Galactic plane of apparently isolated LPTs differs from the scale height of typical pulsars and white dwarfs, which are 350~pc and 300~pc, respectively \citep{2004A&A...425.1009M, 2017ASPC..509..421K}. Magnetars, postulated to be relatively young neutron stars, are concentrated at low Galactic latitudes and have a smaller scale height of only 20-31 pc \citep{MagnetarCatalogue}. Overall, the spatial distribution of LPTs appears more consistently with that of magnetars, supporting a potential association. However, this correlation remains tentative given the small number of detected LPTs and the relative early stage of their study.

Moreover, existing emission theories of pulsars and white dwarfs cannot adequately explain the coherent, high-brightness radio emission observed from LPTs. Magnetars, which power their emission through the decay of magnetic fields, are the more likely progenitors. Several studies support this interpretation: for example, \citet{BeniaminiPopulation} argued that GLEAM-X J1627 is more likely a highly magnetised neutron star than a white dwarf or a rotationally powered pulsar, while \citet{ASKAPJ1935} proposed that a strong magnetic field could power the observed radio bursts. Theoretical models have also been developed to explore how magnetars might generate coherent radio emission brighter than their spin-down luminosity \citep[e.g.,][]{2024MNRAS.533.2133C}.

If LPTs are indeed associated with magnetars, their non-detection in this study may simply reflect the intrinsic rarity of radio-loud magnetars. To-date, only about 30 magnetars are known, including 6 candidates \citep{MagnetarCatalogue}, comprising less than 1\% of the over 4000 catalogued pulsars \citep{ATNFPulsarCatalogue}. Previous studies suggest that high-energy (X-ray emitting) magnetars form at a rate of $\gtrsim 10\%$ relative to the pulsar population \citep{2006csxs.book..547W,2015MNRAS.454..615G,2019MNRAS.487.1426B,2025ApJ...986...88S}. However, their relatively short lifetimes and the fact that not all magnetars emit in the radio band \citep{MagnetarCatalogue}, implies that the observable population of radio-loud magnetars is intrinsically small. Given these constraints, and considering that this study covers only a small section of the Galactic plane, the absence of new LPT detections may simply be a consequence of limited sky coverage and the low probability of encountering a radio-loud magnetar within this region. Alternatively, radio-active LPT states and intermittency in older magnetars may be a rarer occurrence.

We note that several alternative models have been proposed to explain coherent radio emission from long-period sources beyond the magnetar framework, these include early onset of accretion phase in pulsars \citep{2024PASA...41...14A}, strange dwarf stars \citep{2025ApJ...986...98Z}, and compact object binaries in various flavours \citep[e.g.,][]{1983ApJ...274L..71L,2025ApJ...981...34Q}. This search did not result in new detections and therefore cannot directly test these models. Continued observations and future detections will be necessary to test different models and refine our understanding of LPTs.

\subsection{Transient surface density and sensitivity}\label{subsec:sensitivity}
In this section, we compare our transient search with previous searches conducted at ${\sim}1$ GHz radio frequencies. These searches are typically evaluated based on transient surface density (i.e., number of transients detected per square degree) and sensitivity, which represents the faintest transient detectable. We note that the transient surface density is a more appropriate metric for transients that have an isotropic distribution, such as nearby stars and extragalactic sources. Although the fields searched in this work all have low Galactic latitude, the stellar flares found in this work are relatively nearby, most located at distances of around 100~pc, such that variations in Galactic density are not expected to be significant. Therefore, these transients can be considered as isotropically distributed.

Each EMU field covers approximately 37.5 deg$^2$ per scan, %. We estimated the coordinates of corners of each field in J2000 coordinates based on their central positions and converted them to Galactic coordinates. Overlapping fields were then merged using \textsc{shapely} \citep{shapely}
yielding a total effective searched sky area of 750 deg$^2$. The transient surface density $\rho$ can then be calculated as
\begin{equation}
    \rho = \frac{n}{N\cdot\Omega},
\end{equation}
where $n$ is the number of transients found, $N=3614$ is the number of images taken per field during the search, and $\Omega$ is the effective sky area. The transient surface density of this study is therefore $2.21^{+2.60}_{-1.40}\times10^{-6} \rm \, deg^{-2}$.

We adopted equation (11) of \cite{2016MNRAS.458.3506R} to calculate the sensitivity of this work,
\begin{equation}
    {\rm Sensitivity}=6~{\rm RMS}~{\rm exp}\left( \frac{r^2}{2~\rm {HWHM}^2} \right) \rm mJy,
\end{equation}
where RMS is the root-mean-squared of the images, $r$ is the search radius of the beam, HWHM is the half width half-maximum of the primary beam, and the factor of 6 comes from the 6$\sigma$ threshold of the statistical tests. Since we selected candidates only within the primary beam, we set $r = \rm HWHM$. We calculated the median RMS across all images for every observation, obtaining a mean RMS of 1.48 mJy for that SBID. Given that all fields in this study are located near the Galactic plane, we assume the RMS of the search is comparable to this field, yielding a sensitivity of 14.6 mJy.  

\begin{figure*}
    \centering
    \includegraphics[width=0.85\linewidth]{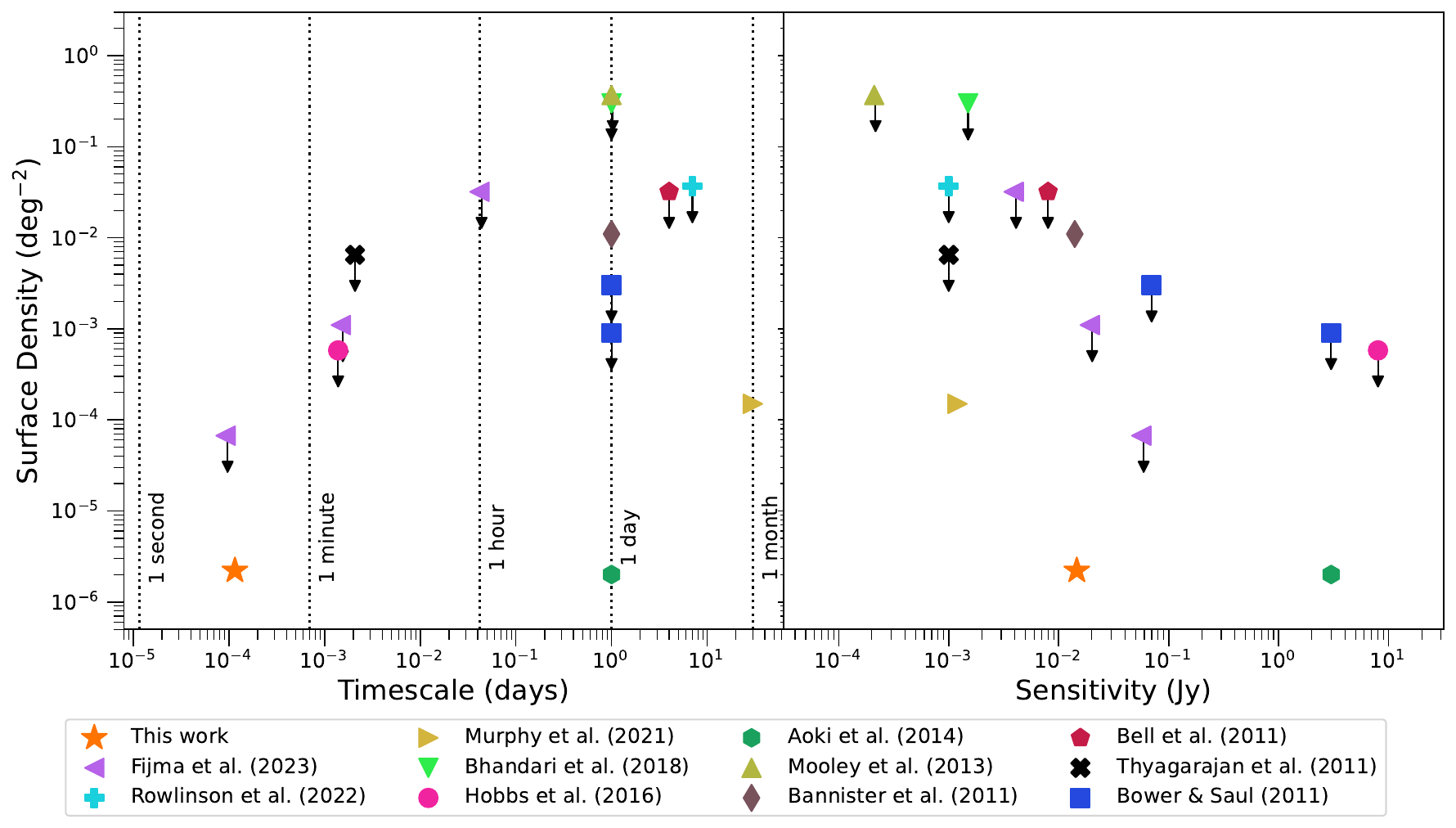}
    \caption{The transient surface density as a function of sampling timescale and sensitivity. This work sets a slightly lower surface density than previous searches of comparable sensitivity and timescale. The reference of different surveys is given in the legend and table \ref{tab:survey_table}. Downward arrows indicate upper limits placed by the surveys.}
    \label{fig:Transient_density}
\end{figure*}

Table \ref{tab:survey_table} lists the timescale, transient surface density, and sensitivity of this study alongside other radio surveys conducted at a similar frequency (${\sim}$1 GHz) for comparison. We also present the results in Figure \ref{fig:Transient_density} to better visualise how this study improves constraints on the transient surface density at mJy sensitivity. \cite{2024MNRAS.528.6985F} used a similar approach, removing persistent sources by creating a deep image of the sky. They conducted a transient search using 7 hours of MeerKAT data at 1.4 GHz, with the shortest timescale of 8 seconds, but did not detect any transients. Using Poisson statistics, they placed an upper limit on the transient surface density of $6.7\times10^{-5} \ \rm deg^{-2}$. In contrast, our work analysed 200 hours of ASKAP data at a similar timescale and detected six transients. This enables us to set a more stringent constraint on the transient surface density at comparable timescales and sensitivity. However, we also note that the transient surface density does not account for the intrinsic nature or spatial distribution of the sources. Coherent, beamed emitters such as pulsars and LPTs are more likely to be detected at greater distances and show strong Galactic latitude dependence, whereas stellar flares are typically nearby and may emit more isotropically. In addition, focusing on densely populated, low-Galactic-latitude fields may lead to bias towards Galactic transients over extragalactic transients. Therefore, while the transient surface density remains a useful comparative metric, a more representative quantity for this work would be the surface density of stellar flares.

\begin{table*}
    \centering
	\caption{The transient surface density and sensitivity of different transient searches conducted at around 1 GHz across various timescales. Figure \ref{fig:Transient_density} visualises these values, mapping the data within the relevant parameter space.}
	\label{tab:survey_table}
	\begin{tabular}{llccc} 
		\hline
		Survey & Central Frequency & Timescale & Sensitivity & Transient surface density \\
             & (GHz) &  & (mJy) & (deg$^{-2}$) \\
		\hline
        \cite{2024MNRAS.528.6985F} & 1.417 & 8 seconds & 56.4 & $<6.7\times10^{-5}$ \\
        This work & 0.9435 & 10 seconds & 14.6 & $2.21\times10^{-6}$ \\
        \cite{2011ApJ...728L..14B} & 1.4 & 60 seconds & 3000 & $<9\times10^{-4}$ \\
        \cite{2016MNRAS.456.3948H} & 0.8635 & 120 seconds & 8000 & $<5.8\times10^{-4}$ \\
        \cite{2024MNRAS.528.6985F} & 1.417 & 128 seconds & 19.2 & $<1.1\times10^{-3}$ \\
        \cite{2011ApJ...742...49T} & 1.4 & 180 seconds & 1.0 & $<6.5\times10^{-3}$ \\
        \cite{2024MNRAS.528.6985F} & 1.417 & 1 hour & 3.9 & $<3.2\times10^{-2}$ \\
        \cite{2014ApJ...781...10A} & 1.42 & 1 day & 3000 & $2\times10^{-6}$ \\
        \cite{2011ApJ...728L..14B} & 1.4 & 1 day & 70 & $<3\times10^{-3}$ \\
        \cite{2018MNRAS.478.1784B} & 1.4 & 1 day & 1.5 & $<3\times10^{-1}$ \\
        \cite{2013ApJ...768..165M} & 1.4 & 1 day & 0.21 & $<3.7\times10^{-1}$ \\
        \cite{2011MNRAS.412..634B} & 0.843 & 1 day & 14 & $1.3\times10^{-2}$ \\
        \cite{2011MNRAS.415....2B} & 1.4 & 4.3 days & 8.0 & $<3.2\times10^{-2}$ \\
        \cite{2022MNRAS.517.2894R} & 1.39 & 1 week & 1.0 & $<3.7\times10^{-2}$ \\
        \cite{VAST} & 0.8875 & 30 days & 1.2 & $1.5\times10^{-4}$ \\
        \hline
	\end{tabular}
\end{table*}

\subsection{Detection rate of stars}
Since all six radio transients found in this study are main-sequence stars, we compare the detection rate of stellar objects with other similar searches in this section.

\cite{VASTER} applied the VASTER pipeline to ASKAP pilot surveys covering 1476 deg$^2$, with most observations lasting 8–10 hours and imaged at a 15-minute timescale. They detected eight stars, yielding a surface density of $5.42 \times 10^{-3}~\rm deg^{-2}$. Based on the 95\% Poisson confidence interval, this corresponds to an expected $4.07^{+3.94}_{-2.32}$ stellar detections over the sky area searched in this work -- consistent with our results. Similarly, \cite{2021MNRAS.502.5438P} conducted a circular polarisation survey using 15-minute ASKAP snapshots at similar frequencies. They identified 33 radio stars across 34,159 deg$^2$, corresponding to a surface density of $9.66 \times 10^{-4}~\rm deg^{-2}$. Our observations are 40 times longer in duration, which we approximate by treating each EMU field as equivalent to 40 separate 15-minute fields. After accounting for this, our scaled surface density becomes $1.54 \times 10^{-4} \ \rm deg^{-2}$, roughly an order of magnitude lower than theirs.

There are several caveats in comparing these studies. First, our 10-second imaging timescale is 90 times shorter than those used in the above works, resulting in an approximately  tenfold decrease in sensitivity. As such, our pipeline is less sensitive to faint stellar flares, particularly those below 2~mJy, which were detectable in previous studies \citep[see, e.g., figure 7 in][]{VASTER}. Second, while \cite{VASTER} focused on high Galactic latitude fields and \cite{2021MNRAS.502.5438P} performed an all-sky survey ($\delta \leq 41^\circ$), our search targeted low Galactic latitudes where stellar density is intrinsically higher due to the concentration of stars in the Galactic thin disk. Lastly, the use of Stokes V in \cite{2021MNRAS.502.5438P} offers a higher sensitivity to stellar objects less affected by noise and background AGN, likely contributing to their higher detection rate. The approximate agreement between our detection rate and that of \cite{VASTER} may reflect the balance between our lower sensitivity and the the higher stellar density in the Galactic plane fields surveyed.

\subsection{Future plans and improvement}

We observe that some transients identified in this study have a $\eta$ value just above the threshold (see Table \ref{tab:all_transients}). In contrast, the candidate list contains a significant number of "noise spikes," which typically account for 30–50\% of all candidates. These noise spikes generally exhibit peak statistics just above the threshold. Therefore, lowering the threshold for $\eta$ while applying a stricter significance level for peak statistics may help uncover more transients while reducing the number of noise spikes. 

Despite the higher threshold in the boxcar convolution step, we noticed many classified "transients" are actually sidelobes sweeping or bright sources drifting into nearby pixels due to the ASKAP calibration error and/or inaccuracies in astrometry. A potential improvement would be to implement a machine learning algorithm to assess the quality of model-subtracted short images around identified candidates, helping to filter out sidelobes. Such an algorithm could also be further trained to identify or classify transients more effectively.

We chose to search the Stokes I data products because not all LPTs display strong circular polarisation, and a total intensity search is therefore the most inclusive approach. Nonetheless, all transients discovered in this work exhibit strong circular polarisation. This result suggests that a dedicated search of Stokes V data could uncover additional sources. Several known LPTs show significant circular polarisation fractions -- about 40 \% for ASKAP J1839–0756 \citep{ASKAPJ1839} and 10–30 \% for GLEAM-X J0704–37 \citep{GLEAM-XJ0704‑37}. A joint search in both Stokes I and V may help mitigate the issue of transients being obscured by nearby bright sources (e.g. UCAC4 129$-$071513 in this study) while retaining sensitivity to weak or unpolarised transients.

All transients identified in this study exhibit relatively low flux densities and pulse widths longer than one hour. On a 10-second timescale, the peak statistics are not sensitive to these transients. The noise in individual images may be comparable to the peak flux density of the transient, resulting in a decrease in S/N$_{\rm peak}$. By performing a similar search on longer timescales, the noise can be averaged out, and the peak flux density of the transient becomes statistically more significant. This averaging could also reduce the processing time per field. Since the duty cycle of LPTs is typically several percent, a minute-timescale imaging search will still be sensitive to LPTs with periods longer than one hour.

Dynamic spectra have been essential in this study for identifying transients. The current pipeline and boxcar convolution are not well suited for detecting broad pulses due to the aforementioned reasons. A potential upgrade to the VASTER pipeline would be to include  dynamic spectra as an output product for each candidate. Additionally, image-based machine learning algorithms could be employed to classify dynamic spectra, reducing the need for manual inspection.

\section{Conclusions}

We conducted a large-scale radio transient search at low Galactic latitudes ($|b|<10\degree$) using the VASTER pipeline with 10-second imaging. This study utilised 200 hours of EMU data observed by the ASKAP telescope, covering 750 deg$^2$ of the sky. The selected fields and processing pipeline were specifically adjusted to search for LPTs.

Although no LPTs were detected, we discovered radio flares classified as transients by our pipeline, originating from six stars. These flares persisted for over an hour, demonstrating the VASTER pipeline’s capability to detect long-timescale transients using second-scale imaging techniques. Notably, at least one of these flares has never been observed in the radio regime before, highlighting the potential for future searches to specifically target such sources. Based on the sensitivity of our study and assuming a beaming geometry typical of canonical pulsars, we estimated a lower bound on the detectable luminosity in this work.

Compared to previous surveys at similar timescales and sensitivities, our study places a more stringent lower limit on transient surface density. The absence of LPT detections may be attributed to their intrinsic rarity and potential bias in the survey pointings' direction. Based on our findings, we propose several improvements and upgrades to the VASTER pipeline and transient search methodologies to enhance the detection of minute-to-hour-timescale transients.

\section*{Acknowledgements}
Y.W.J.L. and T.M. acknowledge funding from the Australian Research Council Discovery Project DP\,220102305. Y.W. acknowledges support through the Australian Research Council grant DP220102305 and FT190100155. M.C. acknowledges support of an Australian Research Council Discovery Early Career Research Award (project number DE220100819) funded by the Australian Government. Parts of this research were conducted by the Australian Research Council Centre of Excellence for Gravitational Wave Discovery (OzGrav), project number CE230100016. The material is based upon work supported by NASA under award numbers 80GSFC21M0002 and 80GSFC24M0006.

This work used resources from the China SKA Regional Centre (CNSRC) \citep{2019NatAs...3.1030A, 2022SCPMA..6529501A}. T.A. acknowledges the support of the Xinjiang Tianchi Talent Program. T.A. and Z.J.X. are supported by the National Key R\&D Program of China (2024YFA1611800), National SKA Program of China (2022SKA0130103) and FAST special funding (NSFC 12041301). This research has made use of the VizieR catalogue access tool, CDS, Strasbourg, France. The original description of the VizieR service was published in \cite{Vizier}. This research has made use of the SIMBAD database, operated at CDS, Strasbourg, France \citep{Simbad}. This research has made use of the data product from TESS \citep{TESS}.

This scientific work uses data obtained from Inyarrimanha Ilgari Bundara, the CSIRO Murchison Radio-astronomy Observatory. We acknowledge the Wajarri Yamaji People as the Traditional Owners and native title holders of the Observatory site. CSIRO’s ASKAP radio telescope is part of the Australia Telescope National Facility (\url{https://ror.org/05qajvd42}). Operation of ASKAP is funded by the Australian Government with support from the National Collaborative Research Infrastructure Strategy. ASKAP uses the resources of the Pawsey Supercomputing Research Centre. Establishment of ASKAP, Inyarrimanha Ilgari Bundara, the CSIRO Murchison Radio-astronomy Observatory and the Pawsey Supercomputing Research Centre are initiatives of the Australian Government, with support from the Government of Western Australia and the Science and Industry Endowment Fund.

This research has made use of numpy \citep{numpy}, matplotlib \citep{matplotlib}, astropy \citep{astropy}, psrqpy \citep{psrqpy}, pygdsm \citep{pygdsm}, casa \citep{Casa}, healpy \citep{healpy}, shapely \citep{shapely}, scipy \citep{scipy}, and astroquery \citep{astroquery}.

%%%%%%%%%%%%%%%%%%%%%%%%%%%%%%%%%%%%%%%%%%%%%%%%%%
\section*{Data Availability}
All of the ASKAP data used in this study are publicly available via CASDA (\url{https://research.csiro.au/casda/}) using the projects codes, coordinates, and SBIDs provided in Table \ref{tab:field_table}.



\bibliographystyle{mnras}
\bibliography{bibtex} % if your bibtex file is called example.bib

%\begin{thebibliography}{99}
%\end{thebibliography}

%%%%%%%%%%%%%%%%%%%%%%%%%%%%%%%%%%%%%%%%%%%%%%%%%%

%%%%%%%%%%%%%%%%% APPENDICES %%%%%%%%%%%%%%%%%%%%%

\appendix

\section{Example light curves of classification pipeline}
\begin{figure*}
    \centering
    \includegraphics[width=1\linewidth]{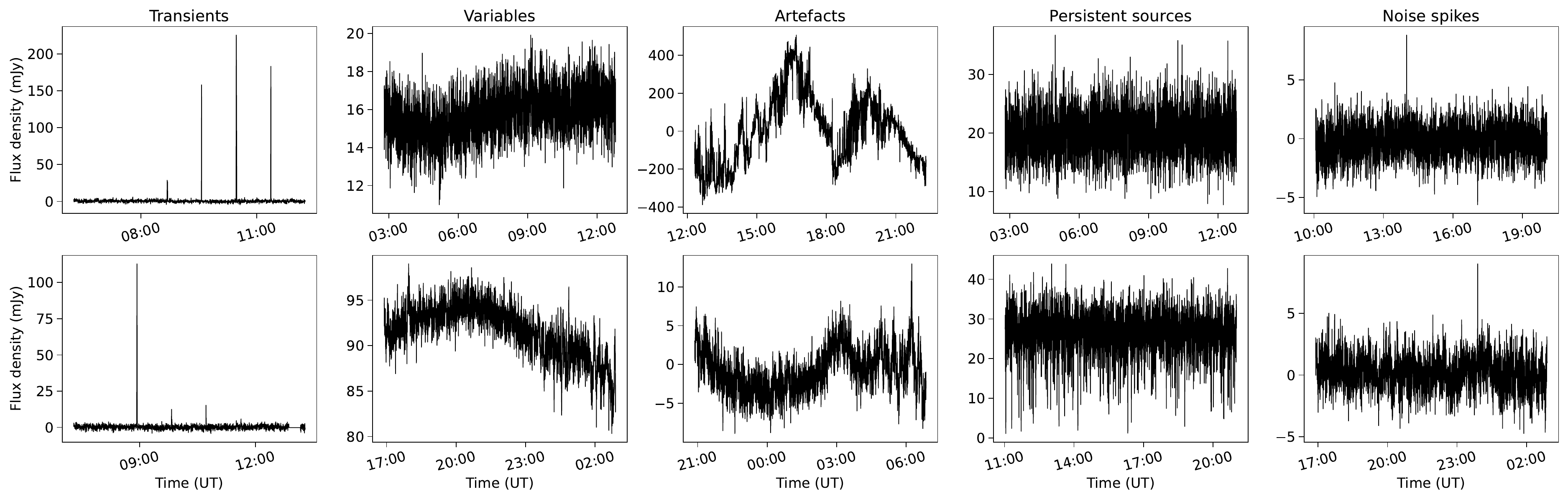}
    \caption{Light curve examples illustrating different types of candidates classified by the candidate classification pipeline described in Section~\ref{sec:classification}. The transient light curves are obtained from 6-hour long observations of ASKAP J1935$+$2148, while the rest are from EMU observations processed in this work.}
    \label{fig:lcexample}
\end{figure*}

In Figure \ref{fig:lcexample}, we present two example light curves for each class of candidates, as classified by the pipeline described in Section~\ref{sec:classification}. The transient light curves correspond to two 6-hour observations of an LPT -- ASKAP J1935$+$2148, which was used to test the classification pipeline. The remaining light curves are genuine candidates identified by the VASTER pipeline in this work.

\section{Additional data on detected stellar flares}
In Table \ref{tab:transients_additional details}, we provide additional information on the stellar flares detected in this work, including peak flux density at 10-second timescale, circular polarisation fraction, full width at half maximum of the pulse, distance to the star, peak emission time in UT, and the z-score ($\sigma_{\rm conv}$) of the light curve convolved with 4-time step boxcar and 64-time step boxcar. The distance of Beta Centauri is adopted from \cite{2016A&A...588A..55P} while others are calculated using the parallax reported from the Gaia DR3 data release \citep{vo:gedr3dist_main, 2021AJ....161..147B}. We only report the z-score of the pulses detected by the pipeline.
\renewcommand{\arraystretch}{1.25}
\begin{table*}
    \centering
    \caption{Additional information on the stellar flares detected in this work. The table provided peak flux density at 10-second time scale, circular polarisation fraction ($\Pi_V$), FWHM of the pulse, distance, peak emission time in UTC, and the z-score convoluted with the smallest and widest boxcar (if applicable). For UCAC4 129-071513, we report the information of both peaks detected in SB55363.}
    \begin{tabular}{c|c|c|c|c|c|c|c}
        \hline
        \multirow{2}{*}{Name} & 10-second peak & \multirow{2}{*}{$\Pi_V$} & \multirow{2}{*}{FWHM (minutes)} & \multirow{2}{*}{Distance (pc)} & \multirow{2}{*}{Emission time (UTC)} & \multicolumn{2}{|c|}{$\sigma_{\rm conv}$} \\
          & flux density (mJy) &  &  &  &  & 4 time steps & 64 time steps \\
        \hline
        Beta Centauri & 6.7 & 91\% & ${>}35$ & $110.6^{+0.5}_{-0.5}$ & 2023-10-21 10:34:11 & - & -\\
        HD 105386 & 3.6 & 86\% & 50 & $427.25^{+4.68}_{-3.84}$ & 2024-04-15 18:44:57 & - & 5.3 \\ 
        Gaia DR3 & \multirow{2}{*}{6.5} & \multirow{2}{*}{82\%} & \multirow{2}{*}{185} & \multirow{2}{*}{$48.02^{+0.06}_{-0.05}$} & \multirow{2}{*}{2023-09-30 05:13:33} & \multirow{2}{*}{-} & \multirow{2}{*}{-}\\[-0.6ex]
        5853594572486546176 & & & & & & & \\
        HD 110244 & 6.6 & 82\% & 133 & $106.68^{+0.98}_{-1.03}$ & 2024-04-14 14:33:16 & - & -\\
        IO Vel & 7.3 & 73\% & 63 & $172.94^{+1.32}_{-1.49}$ & 2023-07-13 10:01:01 & 6.7 & 13.7 \\
        \multirow{2}{*}{UCAC4 129-071513} & 6.7 & 78\% & ${>}53$ & \multirow{2}{*}{$101.51^{+0.20}_{-0.20}$} & 2023-12-13 19:56:19 & - & 7.7 \\
         & 7.8 & 86\% & 56 & & 2023-12-14 04:26:55 & 5.5 & 11.9 \\
        \hline
    \end{tabular}
    \label{tab:transients_additional details}
\end{table*}

\section{Recovery rate of injected transients}
\begin{figure*}
        \centering
        \includegraphics[width=1\linewidth]{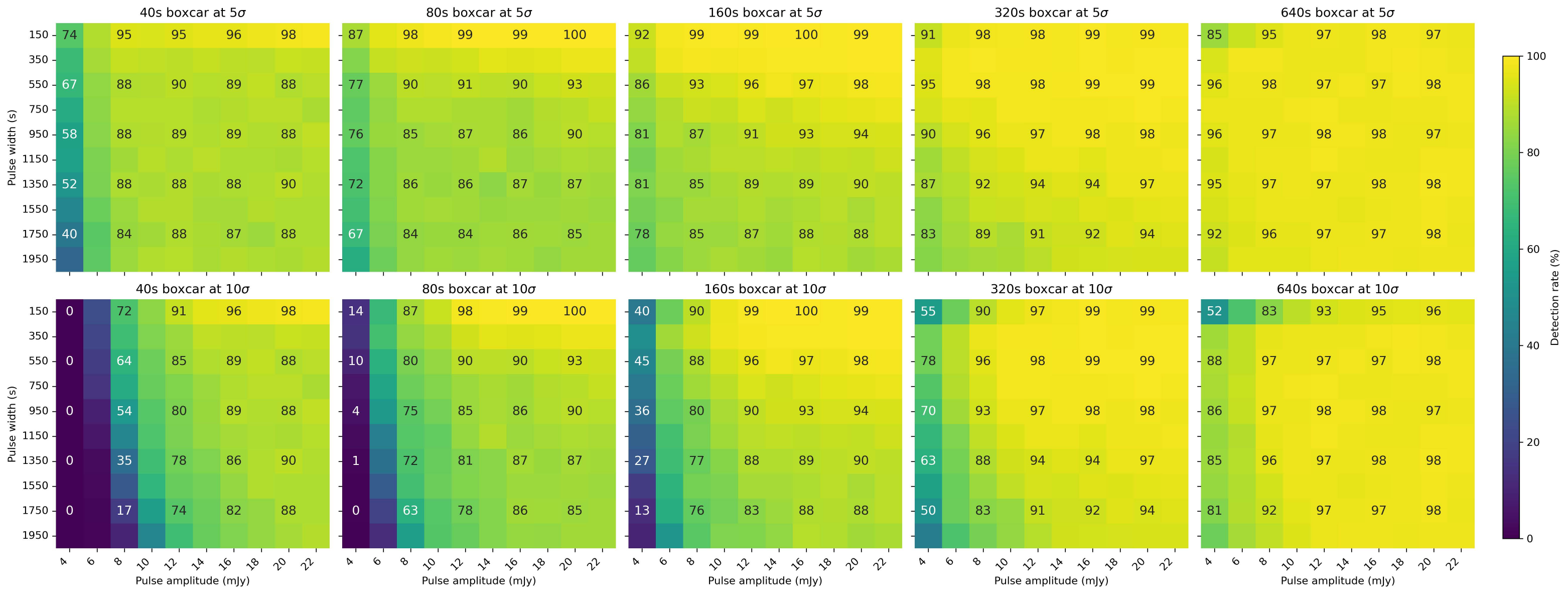}
        \caption{Heat maps showing transient recovery rates for different boxcar widths and $\sigma_{\rm conv}$ thresholds. Injected amplitudes and pulse widths were each divided into 10 bins, forming a 10$\times$10 grid in each map. For every bin, 1000 light curves were injected and the recovery rate was computed. To improve readability, only every other bin is annotated with its recovery rate.}
        \label{fig:recovery_heatmap}
    \end{figure*}

The heat maps of the transient recovery rate using various boxcar widths and $\sigma_{\rm conv}$ thresholds are provided in Figure \ref{fig:recovery_heatmap}. See section \ref{sec:pipeline_veri} for the details of the light curve generation and injection process.

%%%%%%%%%%%%%%%%%%%%%%%%%%%%%%%%%%%%%%%%%%%%%%%%%%

% Don't change these lines
\bsp	% typesetting comment
\label{lastpage}
\end{document}